\documentclass[a4paper,fleqn,usenatbib]{mnras}

\usepackage{amsmath}
\usepackage{amsfonts}
\usepackage{amsbsy}
\usepackage{amssymb}
\usepackage{amsbsy}
\usepackage{makecell}

\usepackage[T1]{fontenc}
\usepackage{ae,aecompl}

\usepackage{graphicx}
\usepackage{gensymb}
\usepackage{float}
\usepackage[utf8]{inputenc}
\usepackage{tensor}
\usepackage{bm}
\usepackage{enumerate}
\usepackage{lastpage}

\newcommand\rmd{{\rm d}}
\newcommand\bb[1]{\mbox{\boldmath{$#1$}}}
\newcommand\grad{\bb{\nabla}}
\newcommand\bcdot{\,\bb{\cdot}\,}

\title[Protostellar disc formation and evolution]{Formation and evolution of protostellar accretion discs. I.~Angular-momentum budget, gravitational self-regulation, and numerical convergence}
\author[W.~Xu and M.~W.~Kunz]{Wenrui Xu$^1$\thanks{Contact e-mail: \href{mailto:wenruix@princeton.edu}{wenruix@princeton.edu}}
 and Matthew W.~Kunz$^{1,2}$\\
$^1$Department of Astrophysical Sciences, Princeton University, Peyton Hall, Princeton, NJ 08544, USA\\
$^2$Princeton Plasma Physics Laboratory, PO Box 451, Princeton, NJ 08543, USA
}

\date{Accepted XXX. Received YYY; in original form ZZZ}

\pubyear{2020}

\begin{document}
\label{firstpage}
\pagerange{\pageref{firstpage}--\pageref{lastpage}}
\maketitle

% Abstract of the paper
\begin{abstract}
We investigate the formation and early evolution of a protostellar disc from a magnetized pre-stellar core using non-ideal magnetohydrodynamic (MHD) simulations including ambipolar diffusion and Ohmic dissipation.
The dynamical contraction of the pre-stellar core ultimately leads to the formation of a first hydrostatic core, after ambipolar diffusion decouples the magnetic field from the predominantly neutral gas.
The hydrostatic core accumulates angular momentum from the infalling material, evolving into a rotationally supported torus; this `first hydrostatic torus' then forms an accreting protostar and a rotationally supported disc.
The disc spreads out by gravitational instability, reaching $\sim$30~au in diameter at $\sim$3~kyr after protostar formation.
The total mass and angular momentum of the protostar-disc system are determined mainly by accretion of gas from an infalling pseudo-disc, which has low specific angular momentum because of magnetic braking; their removal from the protostar-disc system by outflow and disc magnetic braking are negligible, in part because the magnetic field is poorly coupled there.
The redistribution of angular momentum within the protostar-disc system is facilitated mainly by gravitational instability; this allows formation of relatively large discs even when the specific angular momentum of infalling material is low.
We argue that such discs should remain marginally unstable as they grow (with Toomre $Q\sim 1$--$2$), an idea that is broadly consistent with recent observational estimates for Class 0/I discs.
We discuss the numerical convergence of our results, and show that properly treating the inner boundary condition is crucial for achieving convergence at an acceptable computational cost.
\end{abstract}

\begin{keywords}
accretion, accretion discs -- magnetic fields -- MHD -- ISM: clouds -- stars: formation
\end{keywords}

%%%%%%%%%%%%%%%%% BODY OF PAPER %%%%%%%%%%%%%%%%%%
\section{Introduction}

\subsection{Motivation and observations of protostellar discs}

The formation and subsequent evolution of protostellar accretion discs are very important events in the star-formation process. Not only do protostellar discs modulate mass accretion onto young stellar objects and ultimately affect stellar properties such as mass and initial magnetic flux, but also they are directly responsible for establishing the initial conditions for the formation of planets.

The initial conditions for the star- and disc-formation processes have been observed extensively in molecular line emission \citep{Myers1983,Jijina1999,Caselli2002}, infrared absorption \citep{Teixeira2005,Machaieie2017}, and submillimetre dust emission \citep{Ward-Thompson1994,Kirk2005,Konyves2015}, which provide information on the mass, size, temperature, and rotation of pre-stellar cores. The morphology and strength of the magnetic fields have also been inferred via observations of Zeeman splitting \citep[e.g.,][]{Crutcher1999,Falgarone2008} and dust polarization \citep[e.g.,][]{Ward-Thompson2000,Crutcher2004,Girart2006,Maury2018,Auddy2019}. For the later stages of star formation, high-quality observations of circumstellar discs after the dispersal of the pre-stellar envelope (e.g., Class II protostellar discs, or protoplanetary discs) have provided constraints on disc size, mass, and structure \citep[e.g.,][]{Andrews2007,Andrews2013,Pietu2014,Ansdell2018,Andrews2018}.

However, we have much less observational knowledge on how protostellar accretion discs form and grow. Disc formation and growth (if they do grow) are thought to occur mainly during the Class 0/I phase, but these phases are much shorter than either of the pre-stellar or Class II phases. Still, the population of observed Class 0/I discs is growing quickly in recent years and has afforded some interesting insights.
The frequent observation of Class 0 discs suggests that disc formation starts early, but the subsequent evolution is less certain.
A collection of studies measuring the rotational profiles of individual systems suggest large, slowly growing (gas) discs (\citealt{Yen2017} and references within), but surveys in dust continuum tend to find smaller dust discs with either roughly constant \citep{Segura-Cox2018,Andersen2019} or decaying \citep{Tobin2020} size and mass.

The current state of observation makes now a good time for simulations of protostellar disc formation and evolution through the Class 0/I phase, as physical insights obtained from simulations are crucial for interpreting observed trends and understanding the physical picture of disc formation.

\subsection{Physics and simulations of protostellar disc formation and evolution}
\label{subsec:intro_sims}

There is now a large family of simulations that touch on various aspects of protostellar disc formation (see \citealt{Zhao2020review} for a review).
The formation of a rotationally supported protostellar disc is a natural consequence of the pre-stellar core hosting a finite amount of angular momentum.
The source of angular momentum can be inherited from (slow) rotation of the core, injected by turbulence in the molecular cloud \citep{Seifried2012,Joos2013}, or generated during the infall of a core with an asymmetric density profile \citep{Verliat2020}.
In hydrodynamic simulations, conservation of angular momentum produces large, massive discs \citep[e.g.,][]{Bate2018}. In reality, however, the presence of a magnetic field leads to magnetic braking, which can remove most of the initial angular momentum of a pre-stellar core \citep{MP80,BasuMouschovias94}.

The efficacy of magnetic braking depends on the strength and orientation of the initial magnetic flux \citep[e.g.,][]{BasuMouschovias1995,Joos2012,Li2013,Masson2016,Tsukamoto2018,Hirano2020}, as well as the strength of the non-ideal magnetic diffusivity (e.g., \citealt{MP86,Konigl87,Li2011,Dapp2012}), which decouples the magnetic field from the poorly ionized gas. Generally, magnetic braking is weaker when the initial magnetic field is energetically weak, the misalignment between the magnetic field and the rotation axis is significant and/or the magnetic diffusivity is strong. These complex dependencies are compounded by the fact that, amongst the three types of non-ideal effects, Ohmic dissipation and ambipolar diffusion reduce the efficiency of the braking, whereas the Hall effect can either decrease or increase the angular momentum of the infalling gas depending on whether the magnetic field is aligned or anti-aligned with the rotation axis and which species is the dominant current carrier \citep{WardleNg1999,Wardle2004,BraidingWardle2012,Tsukamoto2015,Zhao2020}. In cores that are not initially rotating, the Hall effect can even induce rotation, although realistic values of Hall diffusion in dense cores may not be large enough to produce rotationally supported discs \citep{Krasnopolsky2011}. The exact strength of magnetic diffusivity is also important, and it depends on the level of ionization, which is related to the cosmic ray (CR) ionization rate as well as the abundance and size distribution of dust grains \citep{Zhao2018diffusivity}, where low CR ionization, less dust mass, and larger dust grain size generally increase magnetic diffusivity. The initial dust profile may be inferred from observations, but during the collapse one generally expects this distribution to change as grains coagulate and drift through gas at size-dependent velocities \citep[e.g.,][]{Guilet2020}. However, there is only one protostellar disc formation simulation to date that includes (part of) this kind of active dust evolution \citep{Lebreuilly2020}, and the more common practice is to use a (somewhat arbitrarily) modified dust size distribution with small grains removed \citep[e.g.,][]{Zhao2018}.

Finally, the thermal evolution of the disc may also be important. Relatively realistic modeling of the thermal evolution of the disc in MHD star-formation simulations using radiative transfer is included in a few recent studies \citep[e.g.,][]{Wurster2019}, but it is not so clear whether the outcome of disc formation is significantly different from the more commonly used barotropic equation of state. 

Currently, there are no simulations that incorporate all of the potentially relevant physics mentioned above, but recent simulations are incorporating an increasingly large subset of them. Through these simulations, we gain some  understanding of how the inclusion of various physics affect disc formation. However, a clear physical picture of disc formation is still lacking. More precisely, we do not fully understand what are the main mechanisms regulating the outcome of disc formation, and there is no simple way (other than to run a relatively expensive simulation) to connect reliably and quantitatively the initial conditions of a pre-stellar core to the eventual formation of a rotationally supported accretion disc.

Another issue present in many existing simulations is the difficulty of performing numerically converged, long-term simulations. Here `long-term' means simulating the system until the pre-stellar envelope is largely or fully dispersed, covering most or all of Class 0/I phase.
Such long-term simulations are crucial for understanding observations in Class 0/I systems as well as for connecting with observations and simulations of protoplanetary (Class II) discs.
This difficulty is essentially a problem of computational cost: the physics that need to be modeled, such as non-ideal MHD, often put a very stringent limit on the numerical timestep, and the real-world time of the simulation scales sharply with resolution.
Simulations that directly resolve the protostar show good convergence, but can usually reach no more than 1--2 kyr after protostar formation \citep[e.g.,][]{Machida2019}. The computational cost of long-term simulations can be reduced to an acceptable level by significantly reducing resolution around the protostar, often to $\sim$1~au per cell or more \citep[e.g.,][]{Tomida2017}. However, simulations often fail to achieve good numerical convergence at such low resolution, and behaviour may depend heavily on how the inner boundary (or sink particle) is treated \citep{Machida2014,Vorobyov2019,Hennebelle2020,Wurster2019b}. Additionally, such simulations generally cannot vertically resolve a geometrically thin disc, and it is questionable whether they can accurately capture the transport of angular momentum within the disc (e.g., disc spreading by gravitational instability), which may be an important factor in determining disc size and mass.

\subsection{Our goals}

In this series of studies, our main goals are to understand better the physical picture of protostellar disc formation and eventually to construct a simple, quantitative model that links initial conditions to disc-formation outcomes and allows direct comparison with observations. We plan to work towards these goals with a focus on three aspects. 

First, in terms of simulations, we plan to perform long-term simulations of disc formation and evolution, and parameter surveys covering a sufficiently large parameter space. 
As mentioned in Section \ref{subsec:intro_sims}, keeping the computational cost of such simulations reasonable while maintaining numerical accuracy is a challenging task.
However, we will show in this paper that, with careful, physics-oriented modeling, one can achieve numerical convergence with relatively low resolution.
We also note that, in order to obtain physically correct results, our simulations may need to incorporate a large subset, if not all, of the physical processes discussed in Section \ref{subsec:intro_sims}. In this paper, we will leave out some of them (Hall effect, radiative transfer, dust evolution), but we plan on incorporating some or all of these processes in future work.

In terms of understanding the physical picture of disc formation, we will carefully analyze our simulations with a focus on identifying the relative importance of different physical mechanisms and exploring how the interplay between different mechanisms regulates (and puts physical constraints on) disc evolution.
Physical insights and constraints obtained in this fashion can then be compared against trends seen in observations (e.g., our discussion in \S\ref{subsec:summary_obs}).

Finally, in terms of modeling, we seek to develop either a semi-analytic model for disc formation, or a set of sub-grid models that would allow all important physics to be captured in computationally inexpensive simulations with reduced dimensionality and/or low resolution. This requires physical insights from the previous two efforts, and could involve simplifying the problem by ignoring relatively unimportant physical processes and by capturing the effect of certain physical processes or regulation mechanisms through (analytic or empirical) parametrization.

\subsection{This paper}

In this first paper, we build a basic model including non-ideal MHD effects and use it to simulate the collapse of a pre-stellar core until a few kyr after protostar formation. This covers the formation and early evolution of a protostellar disc. We show that numerical convergence is achieved through a proper treatment of the inner boundary, and we motivate a relatively simple physical picture for disc formation based on our simulation results.
Future papers in this series will focus mainly on incorporating additional physical processes in the model ({\em viz.}~Hall effect, dust evolution), performing parameter surveys to understand any dependencies on physical parameters, and making contact between our simulated protostellar cores and discs and those now being increasingly imaged by mm/sub-mm telescopes.

This paper is organized as follows. In Section \ref{sec:method} we present our simulation setup. In Section \ref{sec:evolution}, we provide an overview of the qualitative evolution of our fiducial 3D simulation. We then discuss the physical origin of the mass and angular-momentum budget of the protostar-disc system in Section \ref{sec:ML} and how redistribution of angular momentum by gravitational instability shapes the disc evolution in Section \ref{sec:disc}. Section \ref{sec:convergence} concerns the numerical convergence of our solution and how that depends on the treatment of the inner boundary. We conclude in Section \ref{sec:summary} with a summary of our physical picture and numerical insights, as well as some discussions of observational hints supporting our physical picture.

%
% METHOD OF SOLUTION
%
\section{Method of solution}\label{sec:method}

We perform a series of two-dimensional (2D; axisymmetric) and three-dimensional (3D) non-ideal MHD simulations in spherical-polar coordinates $(r,\theta,\phi)$ that follow the evolution of a self-gravitating, magnetic, poorly ionized pre-stellar core all the way to a few kyr after the formation of a central protostar. This evolution includes the formation and early evolution of a massive, rotationally supported, protostellar accretion disc. The simulations are performed using the code {\tt Athena++}, equipped with new modules for computing the self-gravitational potential and for solving an equilibrium chemical network. These modules, as well as other minor modifications of the code that are not covered in this section, are detailed in Appendix \ref{A:numerical}.

\subsection{Initial conditions}\label{sec:initial}

We employ the same physical initial conditions for all simulations. The pre-stellar core is initially spherical, with the number density of neutrals given by
\begin{equation}
n_{\rm n}(r) = \max\left\{\frac{n_0}{1+(r/r_0)^2}, n_\infty\right\}.
\label{eq:n}
\end{equation}
We choose an initial central density $n_0=10^4~{\rm cm}^{-3}$ and characteristic core size $r_0=0.1~{\rm pc}$, representative of a typical NH$_3$ core \citep{Jijina1999}. The background density $n_\infty = 500~{\rm cm}^{-3}$, attained in the initial state at  radii $r\ge 0.436~{\rm pc}$, provides a numerical floor that physically represents the density of the ambient molecular gas surrounding the core; this gas is excluded from the calculation of the self-gravitational potential in order to avoid unphysical behavior at large $r$ (see Appendix \ref{A:numerical}). Adopting a mean mass per neutral particle $m_{\rm n} = 2.33m_{\rm p}$ (accounting for molecular hydrogen with 20\% He by number), the total self-gravitating mass enclosed within $r_0$ is then ${\simeq}1.43~{\rm M}_\odot$.
The initial temperature of the gas is set to $T_0 = 10~{\rm K}$, giving an initial isothermal sound speed $c_{{\rm s}0} \equiv (k_{\rm B}T_0/m_{\rm n})^{1/2} \simeq 0.188~{\rm km~s}^{-1}$.

The core is threaded by an initially uniform, vertical magnetic field with strength $B_0=25~\mu{\rm G}$. In the innermost flux tubes of the core, the ratio of self-gravitating mass to magnetic flux is ${\simeq}1.5$ times the critical central value a for collapse, $(3/2)(63G)^{-1/2}$ \citep{ms76}. The core has supercritical mass-to-flux ratio out to a cylindrical radius $R_{\rm sc} \simeq 1.27r_0$.

For $r\leq r_0$, the core is set into uniform rotation with initial angular velocity $\Omega_0 = 0.2c_{\rm s0}/r_0 \simeq 1.22\times10^{-14}~{\rm rad~s}^{-1}$. This corresponds to a ratio between rotational and thermal energy of ${\simeq}0.011$, again consistent with a typical NH$_3$ core \citep{BarrancoGoodman1998}. To avoid excessively high specific angular momentum in the outer part of our domain (where $r/r_0$ can be at most ${\sim}10$), we set the angular velocity to $\Omega(r) = \Omega_0(r/r_0)^{-2}$ for $r>r_0$, corresponding to constant specific angular momentum along radial lines.

We note in passing that many recent simulations of protostar and protostellar disc formation use initial density profiles similar to equation \eqref{eq:n} but with higher central density (often $n_{\rm n}\sim 10^5$--$10^6~{\rm cm}^{-3}$) and smaller characteristic size. This type of initial condition is meant to represent a later phase of pre-stellar collapse.
By contrast, our initial condition represents an earlier phase of the evolution, where most of the region within $R_{\rm sc}$ has approximately uniform density (cf.~figure 8a of \citealt{KM10}). Starting the simulation at a lower density affords enough time for the gas to flatten along magnetic-field lines before the core contracts dynamically. 
Because the magnetic field evolves under near-flux-freezing during the dynamical collapse of a supercritical core, starting at a lower initial density also affords the initially uniform magnetic field time to adjust and evolve consistently with the density profile.

\subsection{Equation of state}\label{sec:EoS}

For physical simplicity and computational expediency, we model the thermal evolution of the system using the following barotropic equation of state (EoS) for the temperature:
\begin{equation}\label{eqn:EoS}
T(n_{\rm n}) = T_0 \left[1+(n_{\rm n}/n_{\rm adia})^{2/3}\right],
\end{equation}
where $n_{\rm adia} = 2\times 10^{11}~{\rm cm}^{-3}$ is value of $n_{\rm n}$ at which the gas transitions smoothly from an isothermal EoS (at lower densities) to a polytropic EoS with index $\gamma=5/3$ mimicking adiabatic evolution (at higher densities). The break-point of the EoS is chosen such that equation \eqref{eqn:EoS} accurately reproduces the dependence of the central core temperature on density obtained in the radiative non-ideal MHD simulation of protostar formation by \citet[][see their figures 4a and 4b]{KM10}, at least up to densities $n_{\rm n}\gtrsim 10^{13}~{\rm cm}^{-3}$, at which point the central temperature becomes ${\gtrsim}100~{\rm K}$ and $\gamma$ approaches $7/5$. We plan to refine this EoS for future work.

\subsection{Non-ideal MHD diffusivities}\label{sec:nonideal}

We include Ohmic dissipation and ambipolar diffusion in our simulations; the Hall effect is neglected. The associated diffusivities are calculated using an equilibrium chemical model that includes electrons, atomic and molecular ions, and a distribution of (neutral, singly negatively charged, and singly positively charged) dust grains divided into 5 bins of different sizes. The chemical model is identical to that detailed in \citet[][\S 4]{KM09}, except that we choose a smaller CR ionization rate of $\zeta_{\rm CR}=10^{-17}~{\rm s}^{-1}$ (similar to other contemporary numerical models of pre-stellar core contraction, e.g., \citealt{Li2011,Marchand16,Tomida15,Zhao2018}) and use a slightly different dust profile.\footnote{At high column densities (${\gtrsim}10^2~{\rm g}~{\rm cm}^{-2}$, see \citealt{UmebayashiNakano1980}), CRs will be shielded by optically thick gas, leading to lower $\zeta_{\rm CR}$ and higher magnetic diffusivities. We ignore this effect in our model and use a constant $\zeta_{\rm CR}$. This should not affect the evolution though, since in our simulation the magnetic field is already barely coupled to the gas (due to ambipolar diffusion) at such high densities.} Namely, we assume spherical dust grains distributed in radius $a$ according to a truncated MRN \citep{mathis77} size distribution, in which the number density of grains with radii between $a$ and $a+\rmd a$ satisfies $\rmd n_{\rm g}/\rmd a \propto a^{-3/5}$ from a minimum grain size $a_{\rm min} = 0.1~\mu{\rm m}$ to a maximum grain size $a_{\rm max} = 0.25~\mu{\rm m}$. The total dust mass is taken to be $1\%$ of the gas mass. The exclusion of dust grains smaller than $0.1~\mu{\rm m}$ is motivated by studies showing that dust drift and coagulation can efficiently remove small dust grains during star formation \citep{Rossi1991,Ossenkopf1993,Ormel2009,Hirashita2012,Guilet2020}.
{The exclusion of small grains also increases ambipolar diffusion and promotes disc formation \citep{Zhao2018}.}
In Appendix \ref{A:diffisivity} we show the abundance of different charged species and the resulting non-ideal diffusivities for our dust model.

\subsection{Computational domain and spatial resolution}

We use a spherical-polar grid with an outer radial boundary at $2^{18}~{\rm au}$ ($1.27~{\rm pc}$, or $12.7r_0$) and a fiducial inner radial boundary at $r_{\rm in}=1~{\rm au}$. The radial ($r$) grid is log-uniform with a fiducial resolution of ${\approx}14$ cells per factor of 2 ($\rmd r/r\approx 0.05$). To reduce computational cost, we only simulate $\theta\in[0,\pi/2]$ with a reflecting boundary condition at the midplane; for our 3D simulations, we only simulate $\phi\in[0,\pi]$ with a periodic boundary condition. The polar ($\theta$) grid is non-uniform with the grid spacing decreasing towards the midplane; the midplane cell size is ${\approx}1/3$ of the polar cell size, and the fiducial resolution is 28 cells in 2D and 24 cells in 3D, resulting in a midplane $\rmd\theta \approx 0.03~{\rm rad}$ (${\sim}2~{\rm deg}$). In 3D, the azimuthal ($\phi$) grid is uniform with a fiducial resolution of 16 cells for $\phi\in[0,\pi]$.

To test the numerical convergence of our simulations, in Section \ref{sec:convergence} we compare simulations covering a range of different $r_{\rm in}$ and resolutions. In particular, we vary $r_{\rm in}$ between $0.25~{\rm au}$ and $2~{\rm au}$ and increase the $\theta$ resolution by up to a factor of 4.

\subsection{Boundary conditions}

For the outer radial boundary, we use an open boundary condition. We note that no visible artifacts appear at that boundary, which is additionally placed far enough away ($12.7r_0$) so as not to influence the evolution of the contracting core. 

For the inner radial boundary, we also use an open boundary condition (density, velocity, and magnetic field in the ghost cells are copied from the adjacent active cell), but with the following modifications. To avoid unphysical mass flow into the active domain from the inner boundary, we cap the radial velocity in the boundary (`ghost') cells at zero. We also directly set the mass flux through a cell interface on the inner boundary to zero if the integrator returns a positive mass flux. Most importantly, we set the flux of vertical angular momentum, $L_z = \int\rmd V \, \rho v_\phi r\sin\theta$, through all cell interfaces on the inner boundary to zero; this prevents any angular momentum from leaving the domain through the inner boundary.\footnote{{Normally, the code computes fluxes across interfaces by solving the Riemann problem and updates the density and momentum based on these fluxes. Our boundary condition for $L_z$ is implemented by zeroing out the radial flux of the $\phi$-component of the momentum on the inner radial boundary before updating the variables using the fluxes.}} 
This condition is motivated by the following. Physically, most mass at $r<r_{\rm in}\sim 1~{\rm au}$ is eventually concentrated in a small protostar, which cannot hold much angular momentum. As a result, the net angular-momentum flux though $r=r_{\rm in}$ should be negligible compared to the angular-momentum flux that an open boundary condition would allow, which is on the order of the mass flux multiplied by the local Keplerian specific angular momentum. This particular angular-momentum-flux boundary condition is seldom, if ever, used in previous studies, but, as we show in \S\ref{sec:convergence}, it is crucial for achieving numerical convergence when $r_{\rm in}$ is not very small. No special treatment for the magnetic field at the inner boundary was necessary, as there does not seem to be any numerical artifacts.

To account for the mass inside the inner radial boundary, we set a point mass at $r=0$ and update its mass according to the mass flux through the inner boundary. The gravitational potential of this point mass accretor adds to that of the self-gravitating gas inside the domain.

\subsection{Definitions}

\textbf{The protostar-disc system}: We use this term to define the region that is supported against gravity (by pressure or rotation), plus the point mass enclosed within the inner radial boundary of the computational domain. When the protostar first forms, this region is the pressure-supported first hydrostatic core; later on, it consists of a protostar (inside the inner boundary) and a disc (which is mainly rotationally supported).

We define the boundary of the protostar-disc system as follows. First, we select all cells above a density threshold of $10^9~{\rm cm}^{-3}$. (For our simulations, the precise choice of this threshold barely affects the results, as there is usually a very large density contrast between the disc and its surrounding envelope or outflow.) We then calculate the radial distribution of (azimuthally averaged) angular momentum $\rmd L_z/\rmd \ln r$ in this dense region. The magnitude of $\rmd L_z/\rmd\ln r$ in the protostar-disc system is much higher than in the surrounding material, and there is a sharp local minimum of $\rmd L_z/\rmd\ln r$ at disc boundary. We use this to define the radial boundary of the disc, $R_{\rm d}$; all cells with $r<R_{\rm d}$ and above the density threshold are counted as part of the protostar-disc system.

\textbf{Protostar, accretor, and disc}: Although the boundary of the protostar-disc system is easy to define, defining the boundary between the protostar and the disc can be tricky. For simplicity, we identify the mass of the point-mass accretor within the inner boundary as the protostar mass and count the remainder of the protostar-disc system as the disc. Note that this definition will count most of the first hydrostatic core as being part of the disc.

\textbf{Pseudo-disc and envelope}: We use the term `pseudo-disc' to refer to the pre-stellar material that is pressure-supported (and thus flattened) along magnetic-field lines but not rotationally supported in the cylindrical-radial direction. The envelope is the lower density region that has yet to reach vertical hydrostatic equilibrium.

\section{The fiducial run: evolution and morphology}
\label{sec:evolution}

\begin{figure*}
    \centering
    \vspace{-4em}
    \hbox{\hspace{2em} \includegraphics[width=\textwidth]{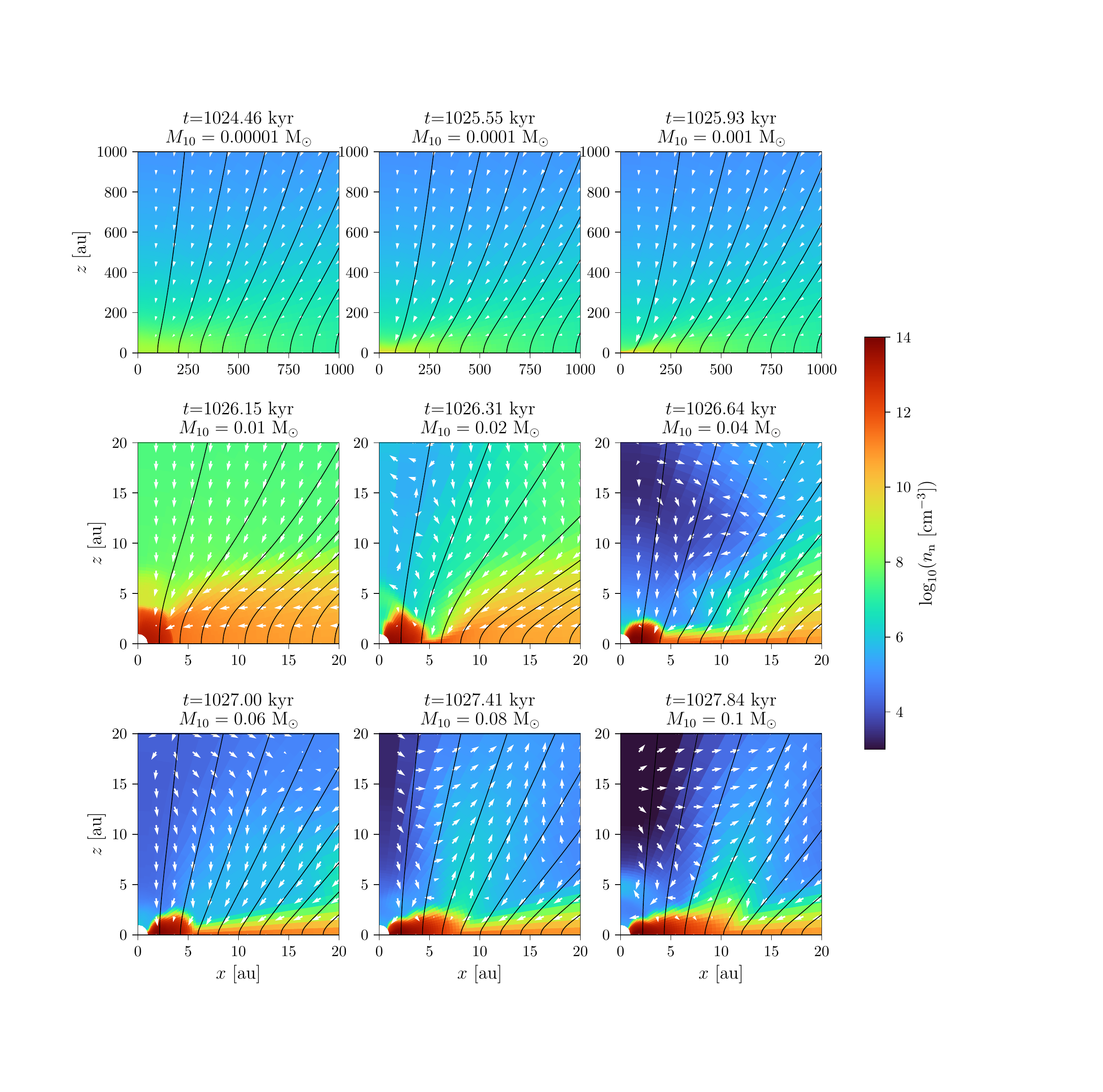}}
    \vspace{-5em}
    \caption{Azimuthally averaged profiles at different epochs from our fiducial 3D simulation. The total mass enclosed within $10~{\rm au}$ (including the accretor mass), $M_{10}$, is provided for reference. Each panel displays the density (colour), magnetic-field lines (black streamlines), and velocity (white quivers). Each row captures a different phase of evolution; from top to bottom: the flattening of the pre-stellar core along magnetic-field lines and the advection-diffusion of the magnetic field into an hourglass shape, the formation of the first hydrostatic core/torus on ${\sim}{\rm au}$ scales, and the spreading of the rotationally supported disc through gravitational instability.}
    \label{fig:snapshot_overview}
\end{figure*}

\begin{figure*}
    \centering
    \includegraphics[width=0.9\textwidth]{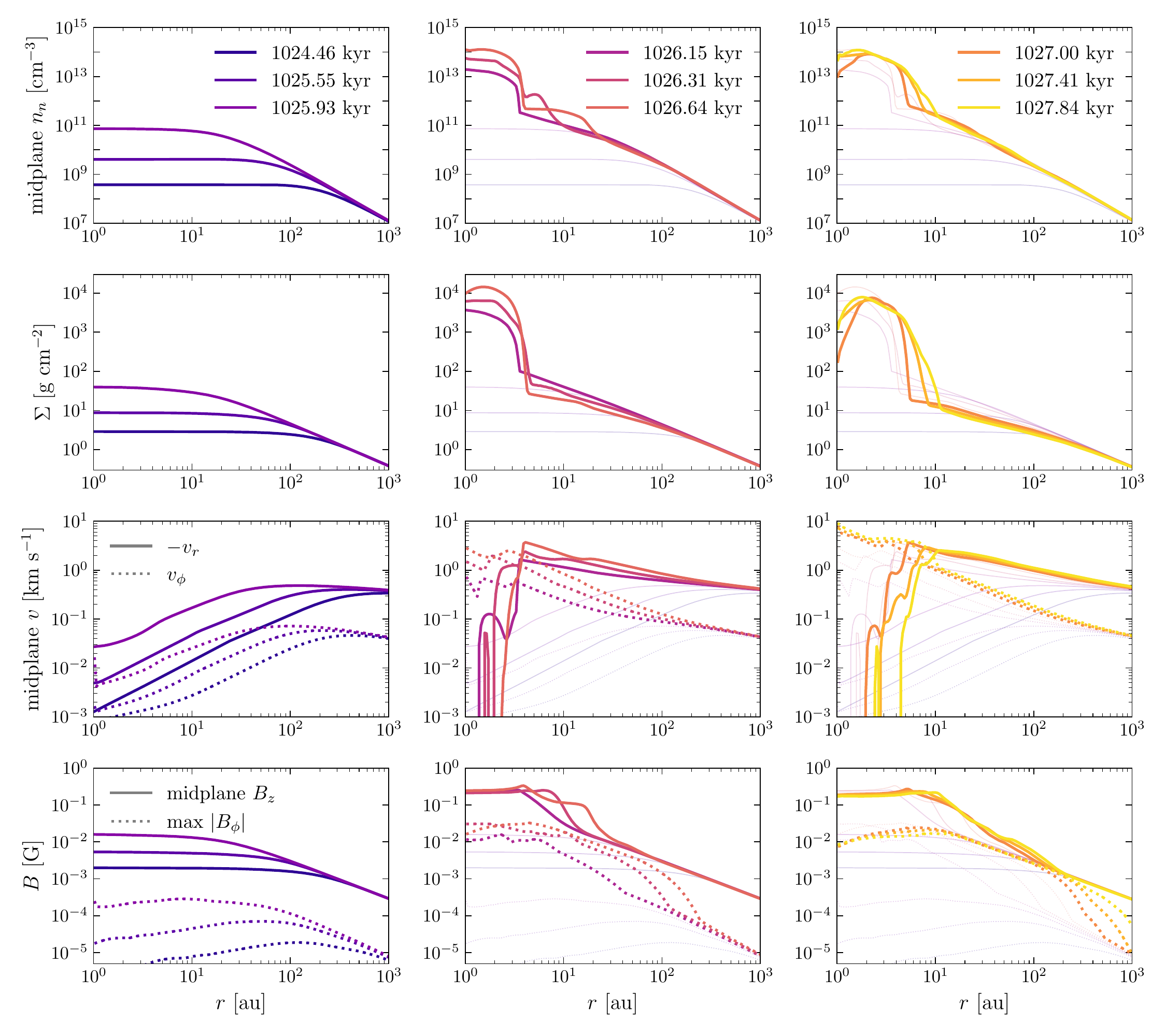}
    \caption{Radial profiles of azimuthally averaged quantities at different epochs from our fiducial 3D simulation. Each column plots three epochs corresponding to a row in Fig.~\ref{fig:snapshot_overview}. To facilitate comparison, earlier epochs are also plotted in later panels using thin lines.}
    \label{fig:curves_overview}
\end{figure*}

\begin{figure}
    \centering
    \vspace{-2em}
    \includegraphics[width=.5\textwidth]{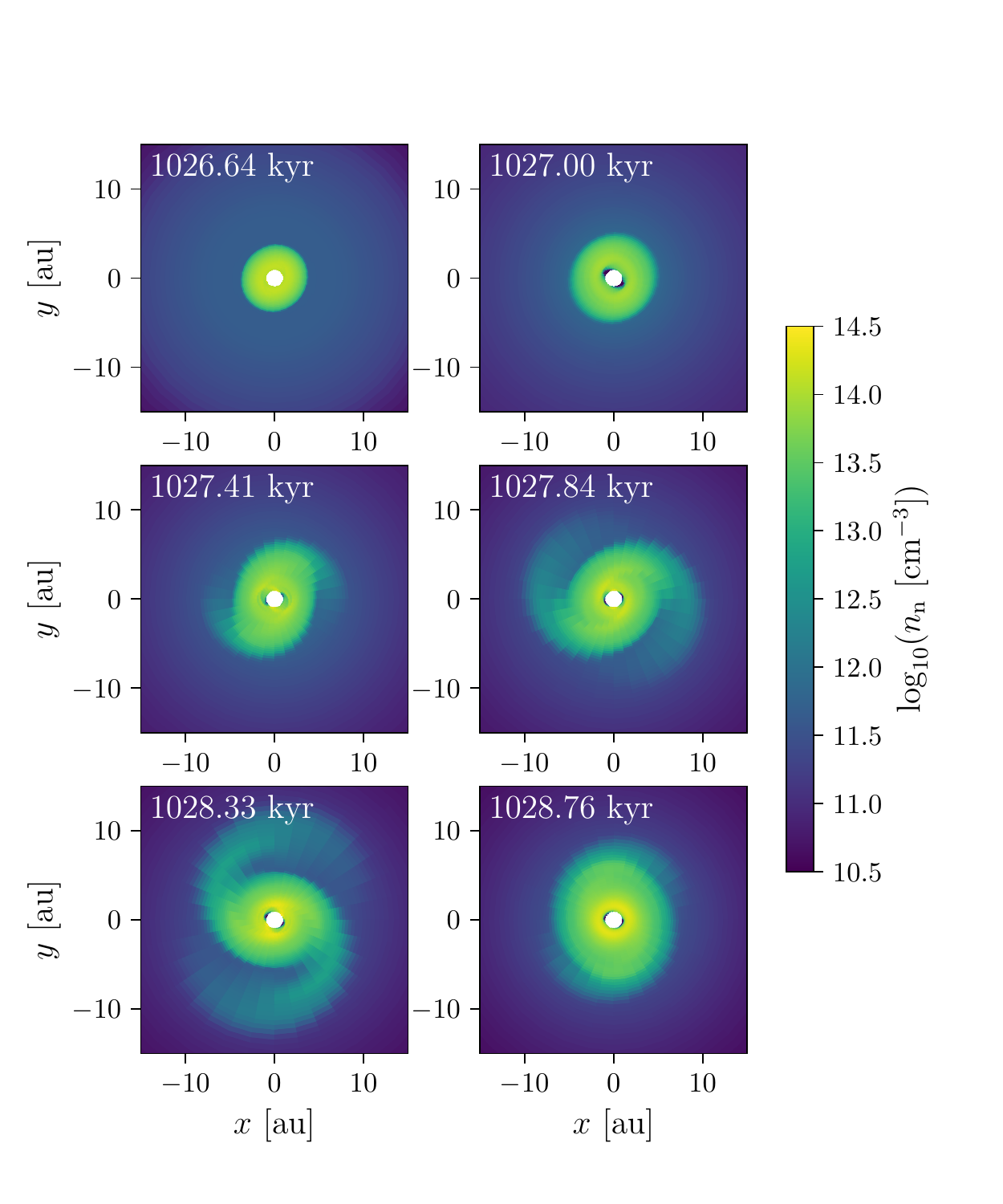}
    \vspace{-2em}
    \caption{Snapshots of midplane density in the fiducial 3D simulation. The first four panels correspond to the last four epochs shown in Fig.~\ref{fig:snapshot_overview}. The initially axisymmetric first hydrostatic torus develops spiral arms due to gravitational instability. These spirals transport angular momentum outward, spreading the disc and aiding protostar accretion. The disc size oscillates as spirals emerge and disperse.}
    \label{fig:snapshot_spirals}
\end{figure}

\begin{figure}
    \centering
    \includegraphics[width=.5\textwidth]{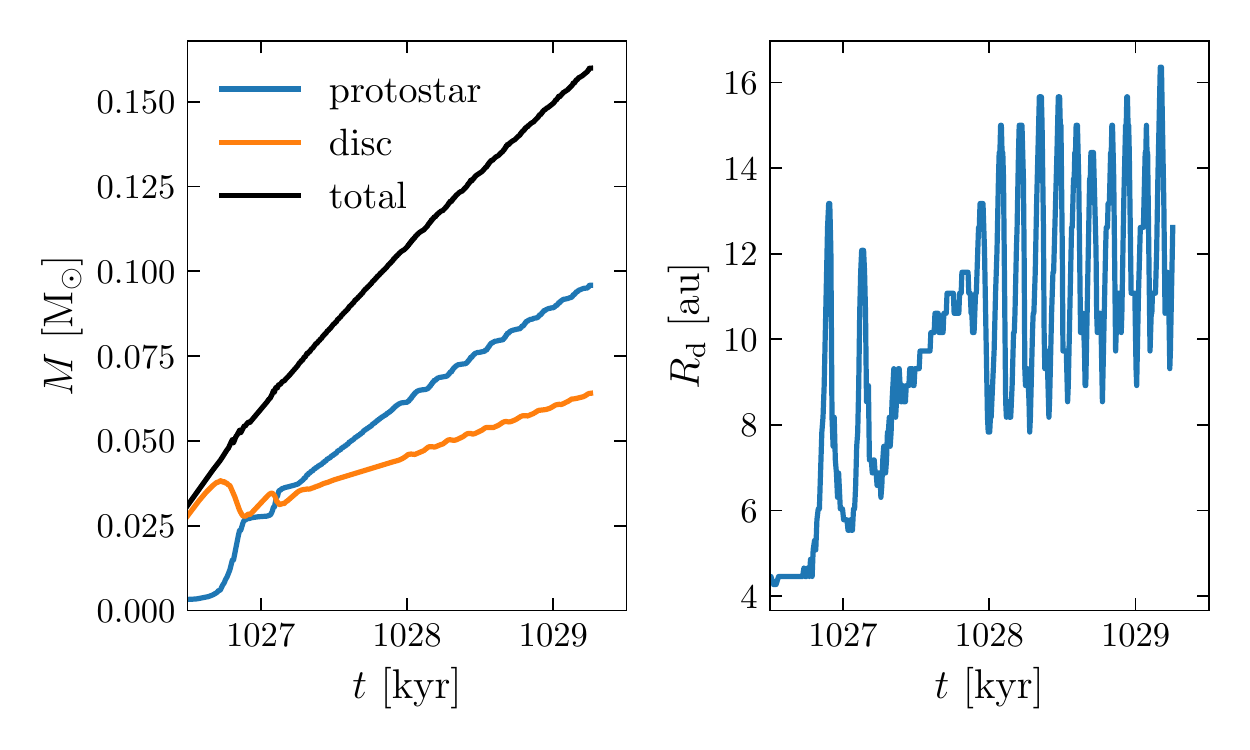}
    \vspace{-2em}
    \caption{Time evolution of protostar mass, disc mass and disc size in the fiducial simulation.}
    \label{fig:disc_mass_size}
\end{figure}

\begin{figure}
    \centering
    \includegraphics[width=.4\textwidth]{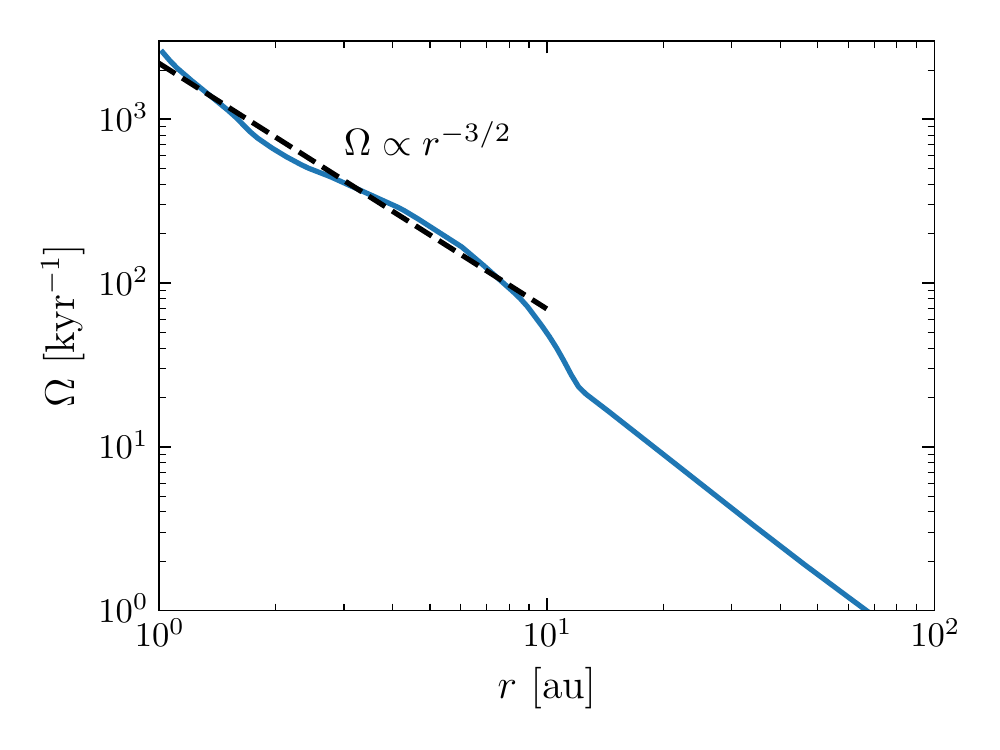}
    \vspace{-1em}
    \caption{Rotation profile of the disc at the end of the fiducial 3D simulation.}
    \label{fig:Omega}
\end{figure}

To begin the discussion of our results, we first provide an overview of the evolution found in our fiducial 3D run. This run covers the evolution from an initially spherical pre-stellar core all the way to a few kyr after protostar formation. Some basic diagnostics of the simulation, including snapshots of the density, velocity, and magnetic field profiles at a number of different epochs, are summarized in Figs~\ref{fig:snapshot_overview} and \ref{fig:curves_overview}. The evolution can be roughly divided into three different phases, each of which are described in the following subsections.

\subsection{Pre-stellar flattening and collapse}

The first phase of the evolution involves the flattening along magnetic-field lines and the dynamical contraction of the pre-stellar core before the formation of the protostar. This covers the first ${\sim}1~{\rm Myr}$ of evolution. The first row of Fig.~\ref{fig:snapshot_overview} and the first column of Fig.~\ref{fig:curves_overview} show snapshots from some representative epochs during this phase. The core is observed to contract slowly under its own self gravity, and the magnetic field is dragged along by the collapsing core to acquire a classic hourglass shape \citep[as in][]{FM93}. The contraction perpendicular to the magnetic field is slower than the flattening along the field lines because of the additional magnetic tension and pressure, and the core becomes significantly flattened into a pseudo-disc.

During this phase, the evolution of the density and magnetic field are largely self-similar (Fig.~\ref{fig:curves_overview}). Due to the initially slow rotation and further removal of angular momentum by magnetic braking, the rotation of the core remains negligible. Evolution during this phase is important for the mass and angular-momentum budget of the protostar-disc system, and will be discussed in more detail in Section \ref{subsec:M} and \ref{subsec:L}.

\subsection{The first hydrostatic core/torus}

Around $1026~{\rm kyr}$, the central density exceeds the break-point of the EoS, the temperature rises, and the first hydrostatic core forms. This time marks the formation of the protostar. The first hydrostatic core is initially approximately spherical (see Fig.~\ref{fig:snapshot_overview}, first panel in second row), matching the evolution found previously in non-rotating simulations. However, the core quickly evolves into a rotationally supported torus due to its acquired angular momentum (Fig.~\ref{fig:snapshot_overview}, second and third panels in second row; at this point the mass within the inner boundary is much less than the torus mass). This feature has also been observed in some previous high-resolution simulations \citep[e.g.,][]{Machida2019} but has not been discussed extensively.

Just prior to the formation of the first hydrostatic core, the magnetic diffusivity increases significantly as a result of the high densities, which cause the gas-phase ions and electrons to be quickly adsorbed onto grain surfaces (see Appendix \ref{A:diffisivity}). This increases the ambipolar resistivity, which facilitates the decoupling of the magnetic field from the predominantly neutral gas \citep{DeschMouschovias2001,TassisMouschovias2007}. As shown in the bottom center panel of Fig.~\ref{fig:curves_overview},
this decoupling leads to a mainly vertical and approximately constant magnetic field in the first hydrostatic core, as well as a pile-up of magnetic flux in the innermost ${\sim}10~{\rm au}$ of the pseudo-disc.\footnote{{This pile-up of magnetic flux results in a  mass-to-flux ratio that locally increases with increasing cylindrical radius, $\rmd\ln(\Sigma/B_z)/\rmd R> 0$. This is a necessary condition for the development of the magnetic interchange instability \citep{LubowSpruit95}, a possibility suggested in the context of magnetic star formation by \citet{LiMcKee96}. However, it is not a sufficient condition: damping due to inefficient coupling between the magnetic field and the predominantly neutral fluid and erasing of the perturbations by the gravitationally driven inflow of the neutrals can prevent the instability from developing \citep{CiolekKoenigl98}. In other words, the minimum timescale of interchange instability, $\tau_{\rm II} = [ (B_z B_R/2\pi\Sigma)\,\rmd\ln (\Sigma/B_z)/\rmd R]^{-1/2}$, must be smaller than both the gravitational timescale, $\tau_{\rm g} = (R/|g|)^{1/2}$, and the ambipolar-diffusion timescale, $\tau_{\rm AD}= \eta^{-1}_{\rm AD} |\rmd\ln B_z/\rmd R|^{-2}$. In agreement with a similar calculation by \citet{TassisMouschovias05}, we find that this never occurs: only at radii where $\tau_{\rm AD}\ll \tau_{\rm II}$ does $\tau_{\rm II}$ approach (from above) and become comparable to $\tau_{\rm g}$. We see no indication of interchange in our simulations. This is likely to be dependent upon the assumed chemistry and initial condition \citep[e.g.,][]{MachidaBasu2020}.}}
The magnetic-field strength saturates at ${\approx}0.2~{\rm G}$, a value similar to that found in previous calculations \citep[e.g.,][]{KM10,Masson2016} and consistent with estimates for the protosolar magnetic field as derived from meteoritic data \citep{LevySonett1978}. The rotationally supported torus is subject to very little magnetic braking because of its high column density and weak coupling to the magnetic field.

\subsection{Gravitational instability and disc spreading}

About $1~{\rm kyr}$ after its formation, the first hydrostatic torus becomes gravitationally unstable, and remains so for the remainder of the simulation. This gravitational instability produces a pair of spiral arms that transport angular momentum outwards (Fig.~\ref{fig:snapshot_spirals}). This angular-momentum transport leads to the formation of a rotationally supported disc that gradually spreads out (Fig.~\ref{fig:snapshot_overview}, last row) and the accretion onto the protostar, which now lies inside the inner boundary.

The evolution of disc size and mass are shown in Fig.~\ref{fig:disc_mass_size}. The total mass of the protostar-disc system increases at an approximately constant rate, and the disc-to-star mass ratio gradually decreases. (In this figure, only the mass interior to the inner boundary is counted towards the protostar mass; the part of the first hydrostatic core/torus in the active domain is counted towards the disc mass.) The disc size increases to ${\approx}15~{\rm au}$ towards the end of the simulation, which occurs ${\approx}3.5~{\rm kyr}$ after protostar formation.
The rotation profile of the disc at the end of the simulation, which is Keplerian out to ${\sim}10$ au, is shown in Fig.~\ref{fig:Omega}.

In Section \ref{sec:disc}, we discuss the effect of gravitational instability and the evolution of disc size and mass in more detail, and provide some analytic estimates concerning the long-term evolution of the disc.

\section{The mass and angular-momentum budget of the protostar-disc system}
\label{sec:ML}

We now turn to a more quantitative discussion of our simulation results. The evolution of the protostellar disc can be addressed by focusing on two questions: How much mass and angular momentum does the protostar-disc system possess (and how are they accumulated)? How is angular momentum redistributed within the protostar-disc system to determine the disc size and mass? These two questions will be addressed in this and the next section, respectively.

\subsection{Contribution to mass and angular-momentum budget from different mechanisms}
\label{subsec:budget}

\begin{figure}
    \centering
    \includegraphics[width=.5\textwidth]{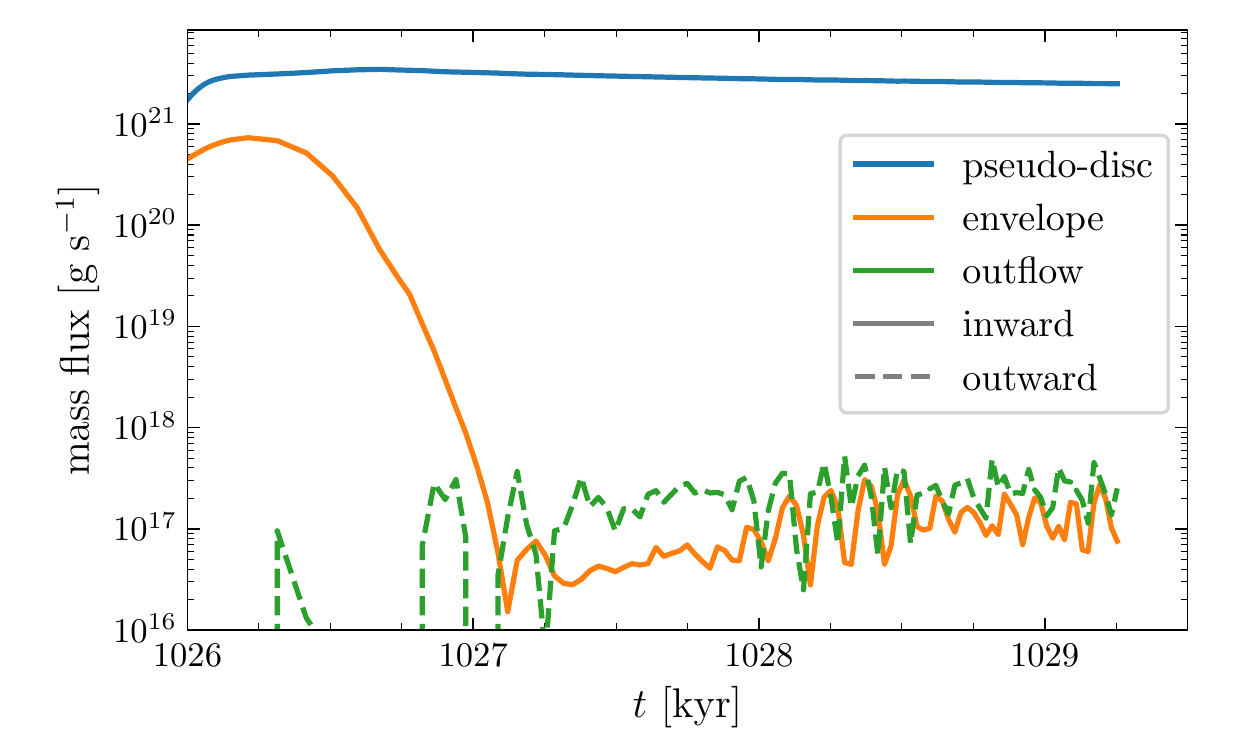}
    \includegraphics[width=.5\textwidth]{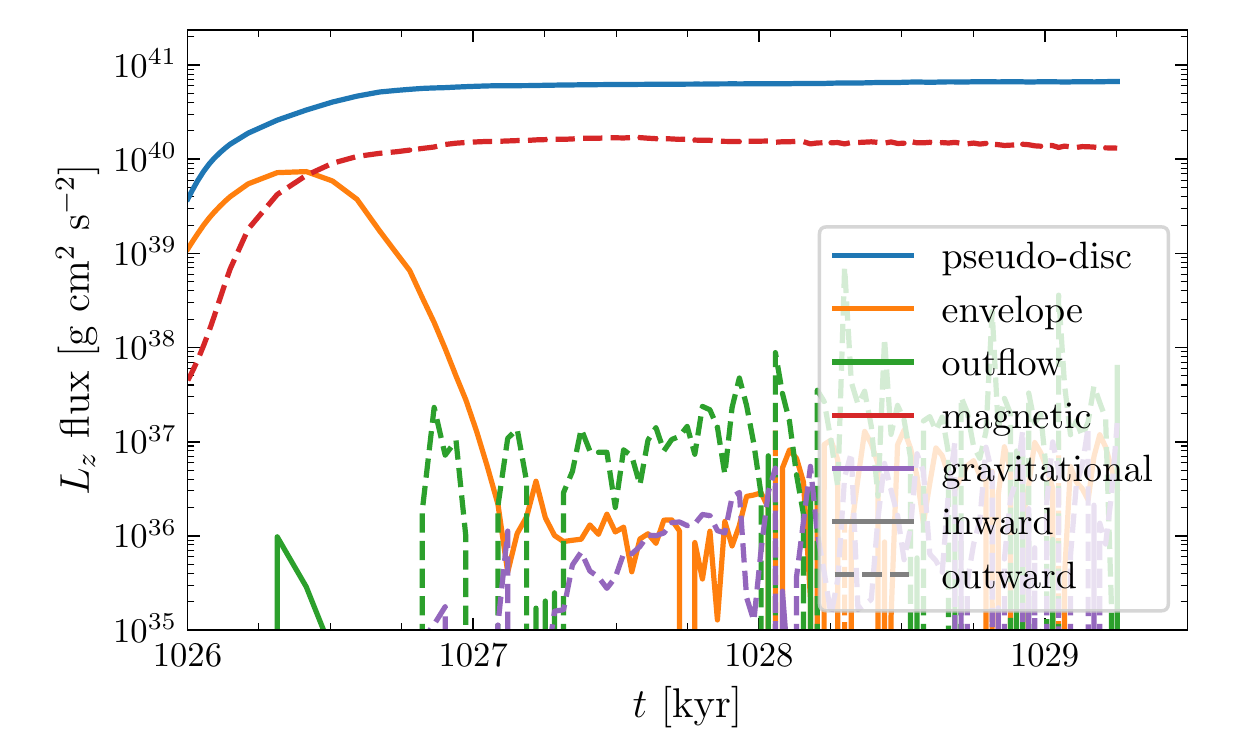}
    \vspace{-1em}
    \caption{Contribution to mass and angular-momentum flux through a 20-au sphere from different mechanisms. Both mass flux and angular-momentum flux are dominated by inflow (accretion) from the pseudo-disc.}
    \label{fig:flux_20au}
\end{figure}

There are multiple mechanisms that contribute to the mass and angular-momentum budget of the protostar-disc system. Inflow (from the pseudo-disc or envelope) and outflow (from the protostar-disc system) injects and removes both mass and angular momentum by advection. Additionally, angular momentum is transported by gravitational and magnetic (Maxwell) stresses. To understand their relative importance, we plot the mass and angular-momentum flux through a $20$-au sphere from each of these mechanisms in Fig.~\ref{fig:flux_20au}. The vast majority of mass and angular momentum within this $20$-au sphere belong to the protostar-disc system, so these fluxes are good approximations for $\rmd M/\rmd t$ and $\rmd L_z/\rmd t$ of the protostar-disc system. For both mass and angular momentum, inflow from the pseudo-disc dominates the flux. The effect of magnetic braking is also nontrivial (resulting in an outward angular-momentum flux), but it is weaker than inflow from the pseduo-disc by more than a factor of $5$. Other mechanisms are all weaker by orders of magnitude. {Specifically, while there are clear indications of outflow in our simulation (see, for example, the last panel in Figure \ref{fig:snapshot_overview}), the mass-loss rate due to outflow is significantly lower than the accretion rate from the pseudo-disc.}

The relative unimportance of outflow in our simulation may appear somewhat surprising, given that several previous studies \citep[e.g.,][]{MachidaHosokawa2013,Tomida2017} found that outflow can remove a significant amount of mass and angular momentum. One important difference between our model and these studies is that we adopt a more realistic profile of magnetic diffusivity, which results in stronger magnetic diffusion inside our dense disc (see Appendix \ref{A:diffisivity}), thereby leading to a weaker toroidal field (Fig.~\ref{fig:curves_overview}) and weaker outflow. Our result is also consistent with recent observation from \citet{Sadavoy2019}, which find no strong toroidal field (at $35$-au resolution) in a protostellar disc survey. There are also some differences in the initial conditions, including our adoption of a stronger, near-critical initial magnetic field, which could lead to substantially weaker outflow as suggested by results from \citet{MachidaHosokawa2013}. Overall, it is important to verify in future studies whether our result is applicable over a broader range of initial parameters. Finally, we note that, although observations of relatively high specific angular momentum in outflows \citep[e.g.,][]{Bjerkeli2016} are sometimes used as evidence that outflows can remove a significant amount of angular momentum, this argument is not grounded firmly unless the mass-loss rate by outflow and the accretion rate of the disc are also known.

For the remainder of this section, we focus only on the pseudo-disc inflow (accretion), as it largely determines the mass and angular-momentum budget of the protostar-disc system. The exclusion of the other mechanisms, which are often highly variable, greatly simplifies the problem.

\subsection{Mass budget: accretion from a free-falling pseudo-disc}
\label{subsec:M}

\begin{figure}
    \centering
    \includegraphics[width=.5\textwidth]{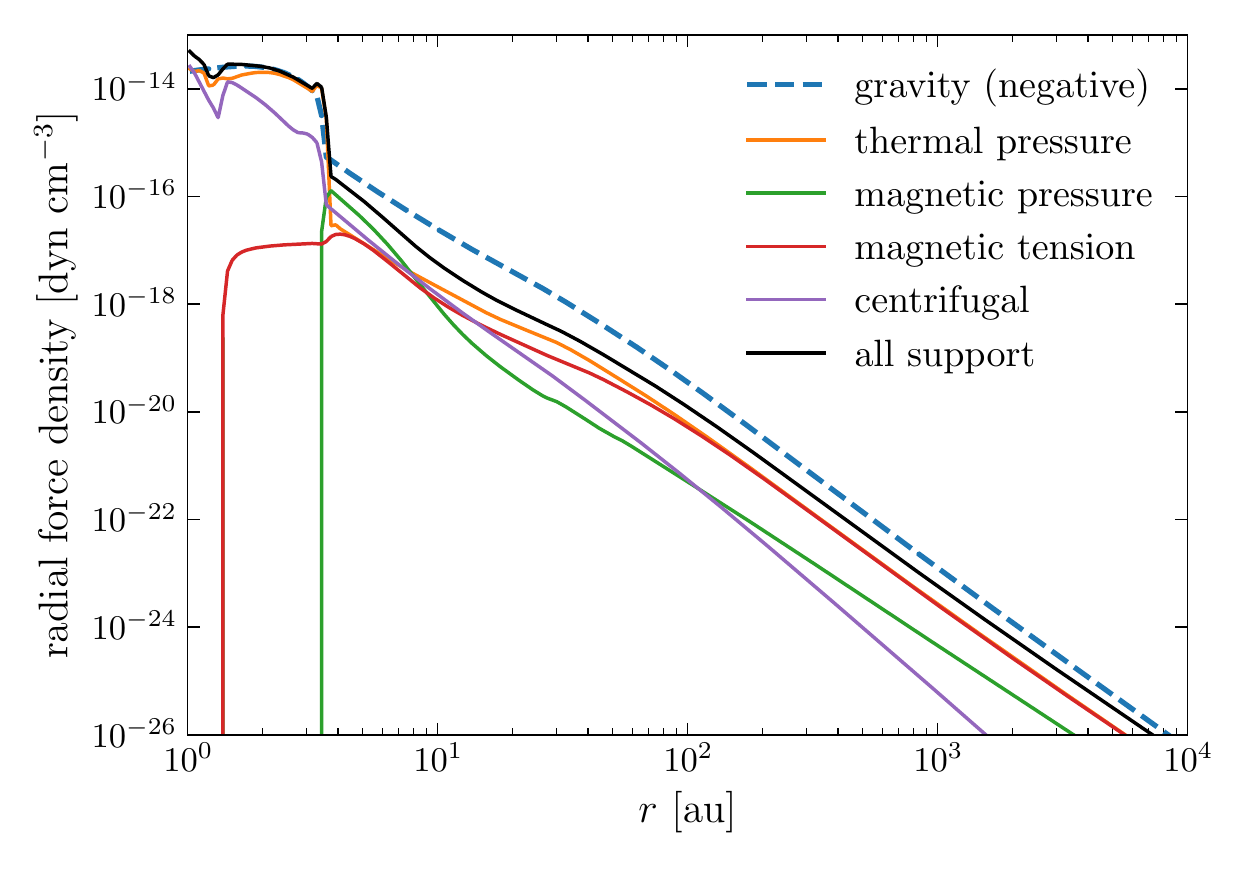}
    \vspace{-2em}
    \caption{Radial force density from gravity and various supporting forces at the midplane, taken from our fiducial simulation when $M_{10}=0.01~{\rm M}_\odot$. Although some supporting forces are of the same order of magnitude as gravity, the sum of them (grey line) remains insufficient to balance gravity in the pseudo-disc, and the pseudo-disc is in near free-fall.}
    \label{fig:radial_force}
\end{figure}

The mass of the protostar-disc system, as discussed in the previous subsection, comes mostly from the pseudo-disc. As Fig.~\ref{fig:flux_20au} shows, the pseudo-disc accretion rate is approximately constant at ${\approx} 3\times 10^{21}~{\rm g}~{\rm s}^{-1}$, or $5\times 10^{-5}~{\rm M}_\odot~{\rm yr}^{-1}$. Here we discuss the origin of this accretion rate by analyzing the pre-stellar collapse phase.

The pre-stellar collapse in our simulation is very similar to the collapse of a thin, slightly supercritical sheet discussed in the semi-analytic model of \citet{Basu1997}. The pseudo-disc shows a flat $\Sigma$ in the central region where thermal pressure smooths out any perturbation and a near-self-similar profile close to $\Sigma\propto 1/r$ outside this central region.
The radial infall of gas in the pseudo-disc is driven by self gravity and is countered primarily by the pressure gradient and magnetic tension (see Fig.~\ref{fig:radial_force}).\footnote{Due to the slow rotation, it is only in the innermost ${\lesssim}20~{\rm au}$ that the centrifugal force becomes dynamically important.}
Although the initial mass-to-flux ratio of the pre-stellar core is just slightly supercritical and $B_z/\Sigma$ does not decrease significantly during most of the infall, the pressure gradient and magnetic forces remain at least a factor of a few smaller than gravity, and the acceleration is ${\gtrsim}50\%$ of free fall in the bulk of pseudo-disc.
Still, this is not to say that these retarding forces are unimportant. On the contrary, the formation of the self-similar column density profile requires a small but nontrivial ratio between these forces and self gravity (similar to the role of pressure in the spherical collapse model of \citealt{Larson1969}), and this sets a characteristic accretion rate that depends only on the isothermal sound speed $c_{\rm s}$ and $G$ (for a given mass-to-flux ratio).
The mass accretion rate calculated from the self-similar model of \citet{Basu1997} is $13c_{\rm s}^3/G \sim 2\times 10^{-5}~{\rm M}_\odot~{\rm yr}^{-1}$, similar to our result.

\subsection{Angular momentum budget: magnetic braking}
\label{subsec:L}

\begin{figure}
    \centering
    \includegraphics[width=.5\textwidth]{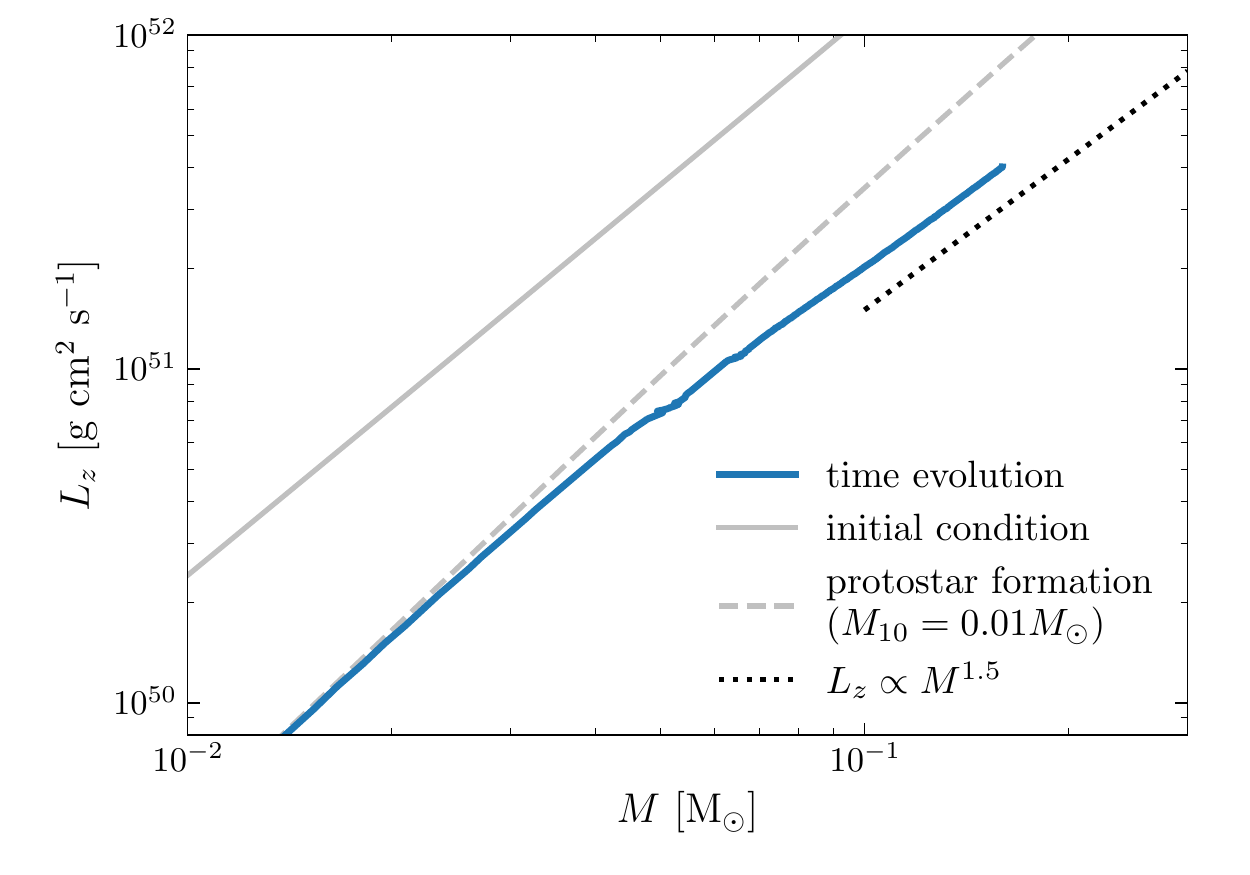}
    \vspace{-2em}
    \caption{Relation between the total mass $M$ and angular momentum $L_z$ of the protostar-disc system (blue line). Here $M$ can be considered as a proxy for time. Above ${\sim}0.1~{\rm M}_\odot$, we see a power-law scaling with slope of ${\approx}1.5$. For reference, we also plot the relation between $M(<r)$ and $L_z(<r)$ at two different epochs: the initial condition (solid grey line) and when $M_{10}=0.01~{\rm M}_\odot$, which is around protostar formation (dashed grey line).}
    \label{fig:ML}
\end{figure}

For the angular-momentum budget, we focus on the relation between the total mass and angular momentum of the protostar-disc system, which is shown in Fig.~\ref{fig:ML}. We see a relatively clean relation between $M$ and $L_z$, which appears to be a power law with slope ${\approx}1.5$ at late times. Unlike in the previous subsection, we are unable to provide a good quantitative explanation for this relation (or its slope) for now, and our discussion will be restricted to understanding the results qualitatively.

The angular-momentum budget is shaped by two main factors: the initial condition, and magnetic braking. In Fig.~\ref{fig:ML} we plot the relation between $M(<r)$ and $L_z(<r)$ for the initial condition; the $M$--$L_z$ relation should be similar (although not identical, since the collapse is not spherical) to this initial $M(<r)$--$L_z(<r)$ relation if there were no magnetic braking. The actual $M$--$L_z$ relation gives much lower $L_z$, showing that magnetic braking has reduced angular momentum by about an order of magnitude.

We can also look at when and how magnetic braking happens. In Fig.~\ref{fig:ML} we also plot the $M(<r)$--$L_z(<r)$ relation at $M_{10}=0.01~{\rm M}_\odot$, which is around the epoch of protostar formation. This curve has a slope of ${\approx}2$, consistent with that of a flattened core with uniform column density and rotation. This suggests that the bulk of the pre-stellar core has undergone a similar amount of magnetic braking. This is reasonable since most of magnetic braking takes place at the beginning of the collapse phase; once the collapse becomes dynamical (with near free-fall velocity), braking is generally slower than the timescale of collapse and angular momentum is approximately conserved.
The $M(<r)$--$L_z(<r)$ relation at $M_{10}=0.01~{\rm M}_\odot$ is close to the  $M$--$L_z$ relation of the protostar-disc system for $M\lesssim 0.05~{\rm M}_\odot$, but the slope of the protostar-disc $M$--$L_z$ relation becomes less steep at later time, suggesting that magnetic braking continues to decrease angular momentum after protostar formation.

It is also worth pointing out why magnetic braking can be important in the pseudo-disc but is always unimportant in the protostellar disc.
This is mainly because the density and column density in the protostellar disc are much higher than in the pseudo-disc (for example, see Figs \ref{fig:snapshot_overview} and \ref{fig:curves_overview}), which makes the magnetic field less well coupled and less dynamically important. The distinction between a dense, non-magnetized disc and a thin, magnetized pseudo-disc has also been observed, for instance, in \citet{Masson2016}.

\section{Evolution of disc size and mass}
\label{sec:disc}
In the previous section we discussed the mass and angular-momentum budget of the protostar-disc system as a whole. Now we will focus on the redistribution of angular momentum within the protostar-disc system, which determines the evolution of disc size and mass.

\subsection{Angular momentum transport by gravitational instability}
\label{subsec:transport}

\begin{figure}
    \centering
    \includegraphics[width=.5\textwidth]{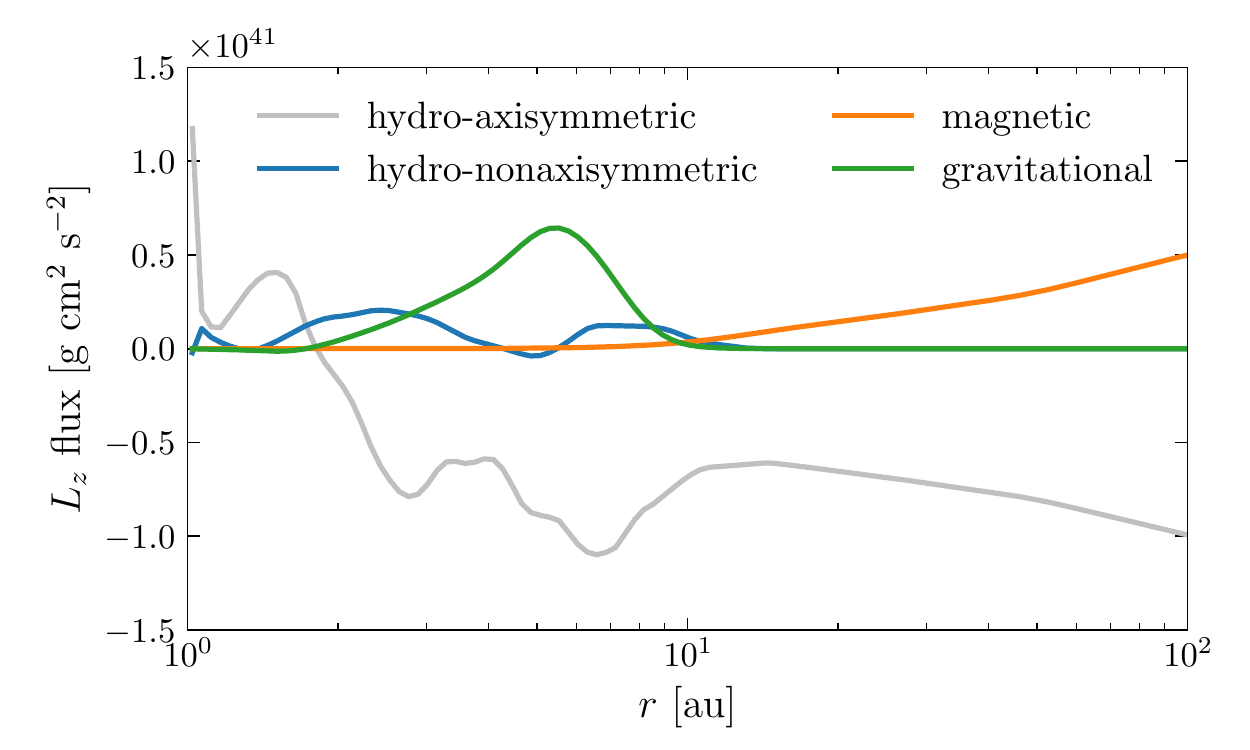}
    \vspace{-1em}
    \caption{Angular-momentum flux due to different mechanisms through spherical shells at different radii, averaged for $t>1028~{\rm kyr}$. The axisymmetric hydrodynamic flux (grey) corresponds to advection with the mass accretion and outflow and does not contribute to the radial transport of angular momentum with respect to the gas. Inside the disc, the transport of angular momentum with respect to the gas is dominated by gravitational instability, which produces the non-axisymmetric hydrodynamic flux (blue) and gravitational stress (green).}
    \label{fig:l_flux_t_avg}
\end{figure}

Soon after the formation of the first hydrostatic core/torus, the dense torus becomes gravitationally unstable, and remains so for the rest of our fiducial simulation. Gravitational instability plays an important role for disc growth, as it transports angular momentum outward and spreads the disc. To see this effect clearly, we plot the contribution from different stresses to the angular-momentum flux through spheres at different radii during disc expansion in Fig.~\ref{fig:l_flux_t_avg}. 
The angular-momentum flux is the sum of hydrodynamic advection (Reynolds stress $\rho v_r v_\phi$), magnetic braking (Maxwell stress $-B_rB_\phi/4\pi$), and gravitational torque (gravitational stress $g_rg_\phi/4\pi G$).
The contribution from hydrodynamic advection can be further decomposed into an axisymmetric component, defined as the product of azimuthally averaged specific angular momentum and mass flux, and a non-axisymmetric component.
Among these four terms, the \mbox{axi}symmetric hydrodynamic flux corresponds to advection due to mass accretion, and does not change the specific angular momentum at given radius. The Maxwell stress is weak inside the disc.
The two remaining terms, the non-axisymmetric hydrodynamic flux and the gravitational stress, are mainly due to gravitational instability and dominate angular-momentum transport within the disc. The sum of these two terms first increases then decreases back to zero, suggesting that gravitational instability takes angular momentum from the inner part of the disc and deposits it in the outer part of the disc. This transport of angular momentum leads to disc spreading, and is the main mechanism that allows protostar accretion when outflow and magnetic braking in the disc are weak.

One important consequence of this gravitational disc spreading is that the disc can now have significantly higher specific angular momentum than the pseudo-disc inflow that feeds the protostar-disc system. For example, Fig.~\ref{fig:Omega} shows a significant jump in $\Omega$ between the disc and the pseudo-disc; similar jumps in $v_\phi$ are also visible in the middle right panel of Fig.~\ref{fig:curves_overview}. This makes having relatively large discs possible even when the pseudo-disc has low specific angular momentum due to magnetic braking.

\subsection{Gravitational self-regulation}
\label{subsec:self_regulation}

The gravitational stability of a geometrically thin isothermal disc can be described by the Toomre $Q$ parameter, defined as
\begin{equation}
Q \equiv \frac{c_{\rm s}\kappa}{\pi G\Sigma}.
\end{equation}
Here $\kappa$ is the epicyclic frequency. The disc is unstable to axisymmetric perturbations for $Q<1$; for non-axisymmetric perturbations the stability threshold increases to $Q\sim 2$. Simulations generally find faster angular-momentum transport by gravitational instability for smaller values of $Q$ (stronger gravitational instability). For values of $Q$ close to (or slightly below) unity, the timescale of angular-momentum transport can be comparable to the orbital timescale (effective $\alpha$ is $\sim 1$), making gravitational instability a very efficient angular-momentum transport mechanism (see \citealt{KratterLodato2016} for a review).

The steep dependence of the angular-momentum transport rate on disc column density (through $Q$) implies a self-regulation mechanism. Higher column density (lower $Q$) leads to faster angular-momentum transport and disc spreading (as well as faster protostar accretion).
Thus there exists a stable equilibrium where disc spreading (which tends to decrease disc column density) balances accretion from the pseudo-disc (which tends to increase disc column density).
Maintaining this kind of self-regulated disc spreading requires a small but nonzero effective $\alpha$. Therefore, most of the disc should be marginally gravitationally unstable with $1\lesssim Q\lesssim 2$.

\begin{figure}
    \centering
    \includegraphics[width=.5\textwidth]{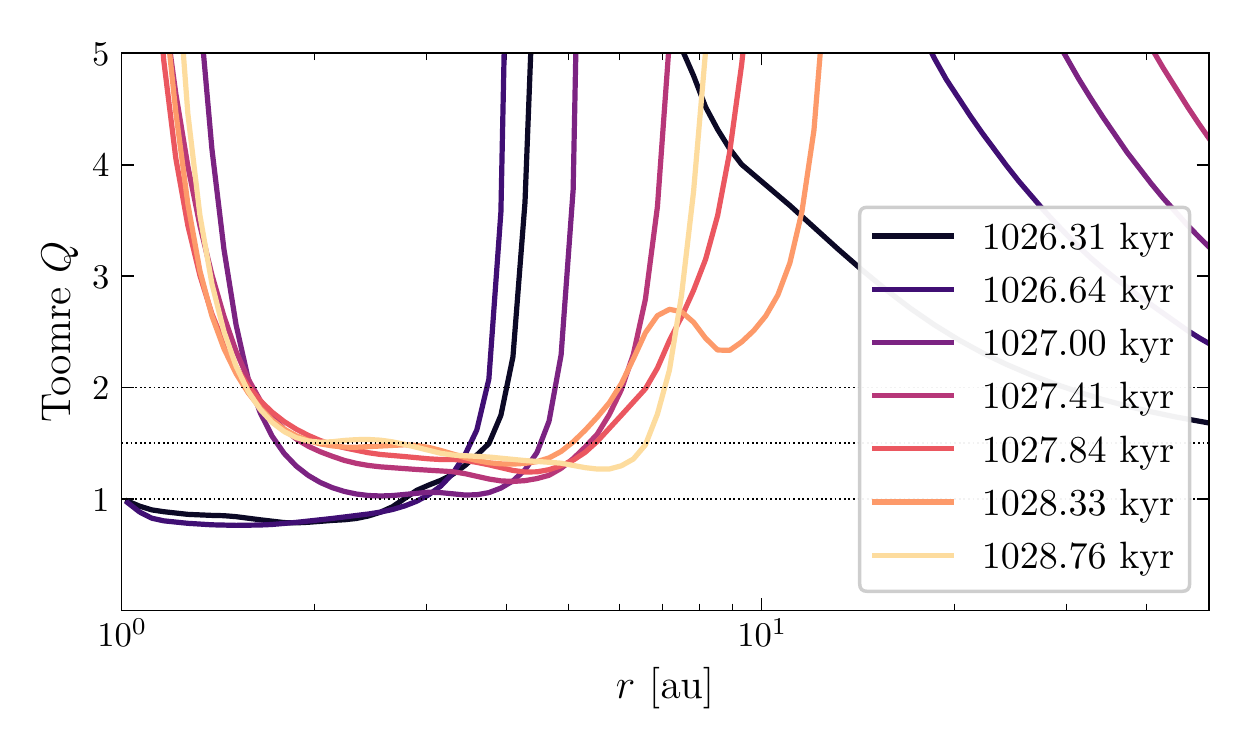}
    \vspace{-1em}
    \caption{Toomre $Q$ (computed using azimuthally averaged profiles) in the fiducial 3D simulation. Each curve corresponds to a panel in Fig.~\ref{fig:snapshot_spirals}. The gravitationally unstable disc generally has $Q$ between 1 and 2.}
    \label{fig:disc_Q}
\end{figure}
\begin{figure}
    \centering
    \includegraphics[width=.5\textwidth]{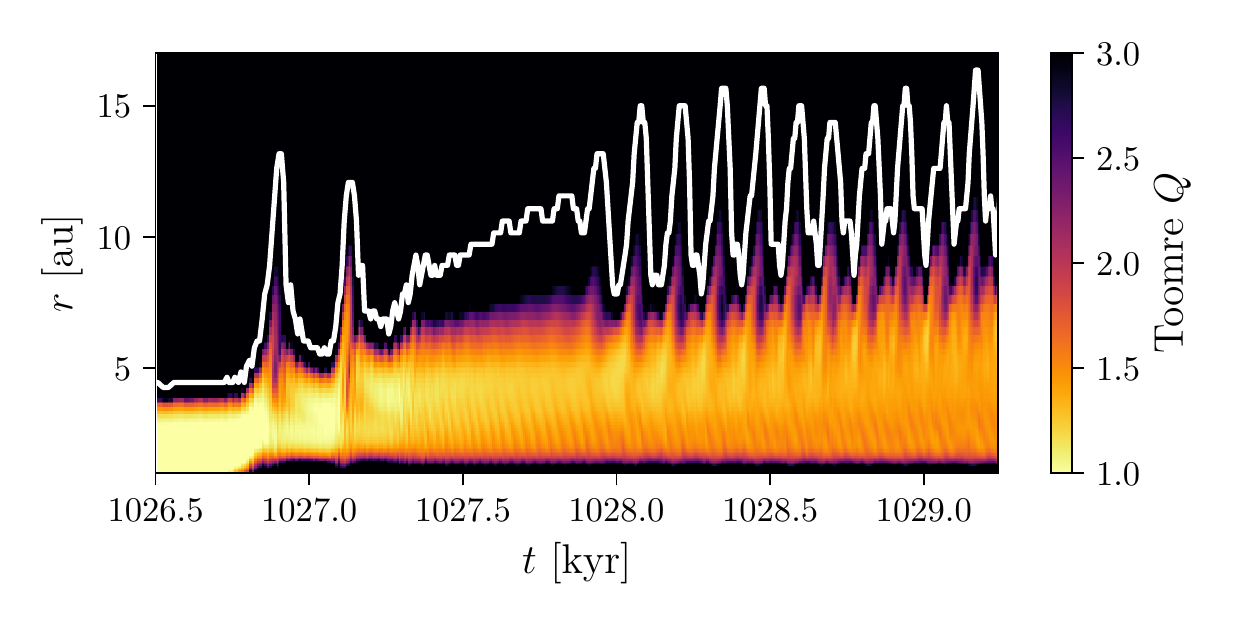}
    \vspace{-1em}
    \caption{Similar to Fig.~\ref{fig:disc_Q} but shows time evolution of the Toomre $Q$ profile. The disc size is also plotted (white curve) for reference.}
    \label{fig:disc_Q_and_Rd}
\end{figure}

This idea of gravitational self-regulation, which implies a marginally unstable disc, is confirmed in Figs \ref{fig:disc_Q} and \ref{fig:disc_Q_and_Rd}, which show that the value of $Q$ does indeed stay between 1 and 2 for most of the disc during its spreading.\footnote{Toomre $Q$ becomes a good description of gravitational instability only at later times ($t\gtrsim 1027~{\rm kyr}$ in our simulation), when the disc is relatively thin and less massive than the protostar. When the first hydrostatic torus first becomes gravitationally unstable (the first two curves in Fig.~\ref{fig:disc_Q}), most of the mass is still in the torus (with the central point mass being negligible), and the stability of the torus is not directly determined by $Q$ \citep[see][]{TohlineHachisu1990}.}
Similar results have also been observed in several other studies
(e.g., \citealt{VorobyovBasu2007}; also see \citealt{LaughlinBodenheimer1994,Tomida2017}).

We also comment that, historically, gravitational instability is often considered to regulate mainly the disc temperature by balancing radiative cooling with heating through spiral shocks and turbulence. This thermal self-regulation, first proposed by \citet{Paczynski1978}, is missing in our current model due to the adoption of a barotropic EoS. In reality, gravitational instability controls disc evolution through both spreading (angular-momentum transport) and heating, and they should be equally important. When the disc is in steady-state, the rate of heating and angular-momentum transport are comparable since both are proportional to the rate at which gravitational instability extracts energy from differential rotation \citep{Gammie2001}.

\subsection{Predicting disc evolution}
\label{subsec:disc_size}

Using the idea of gravitational self-regulation, for given total mass $M$ and angular momentum $L_z$ of the protostar-disc system, one can obtain a robust estimate for the surface density profile, mass, and size of the disc if the thermal profile of the disc is known. Here we perform this estimate for an isothermal disc (with sound speed $c_{\rm s0}$) as an example.

Since we know most of the disc should be marginally stable, we may assume that the whole disc has constant $Q=Q_0$.
For simplicity, we also assume that the disc is not very massive, so $M_\star\approx M$ and $\kappa(R)\approx \sqrt{GM/R^3}$. 
The surface density profile of the disc is then
\begin{equation}
\Sigma(R) \approx \frac{c_{\rm s0}}{\pi G Q_0}\sqrt{\frac{GM}{R^3}}.
\label{eq:Sigma_est}
\end{equation}
This $\Sigma\propto R^{-3/2}$ profile is also observed in the gravitationally regulated disc evolution of \citet{VorobyovBasu2007}.
For disc size $R_{\rm d}$ and disc inner boundary $R_{\rm d,\rm in} \ll R_{\rm d}$, the mass and angular momentum of the disc are then
\begin{align}
 M_{\rm d} &\approx  \int_0^{R_{\rm d}} 2\pi R\rmd R \, \Sigma(R) \approx \frac{4c_{\rm s0}M^{1/2}R_{\rm d}^{1/2}}{G^{1/2}Q_0}, \\*
 L_{\rm d} &\approx \int_0^{R_{\rm d}} 2\pi R\rmd R \, \Sigma(R)\sqrt{GM R} \approx \frac{2c_{\rm s0}M R_{\rm d}}{Q_0}.
\end{align}
Since most angular momentum of the protostar-disc system is within the disc, we have $L_{\rm d}\approx L_z$, which gives
\begin{align}
R_{\rm d} &\approx \frac{L_zQ_0}{2c_{\rm s0}M},\\
M_{\rm d} &\approx \left(\frac{8c_{\rm s0}L_z}{GQ_0}\right)^{1/2}.
\end{align}
Using the $M$--$L_z$ relation of our fiducial simulation (extrapolated assuming a power-law slope of 1.5), the above estimate gives disc mass ${\approx}0.15~M_\odot$ and size ${\approx}80~{\rm au}$ when the protostar reaches ${\approx}1~{\rm M}_\odot$.

In reality, a marginally gravitationally unstable disc is often optically thick and not isothermal. The estimate above is generally an upper limit of the disc size due to gravitational spreading.
%, and may or may not be a good approximation of the actual disc size depending on the actual temperature profile.
Moreover, if the temperature scales too steeply with radius (which is the case for the particular barotropic EoS we use in this work, but is generally not true for a disc with realistic cooling), most of disc mass and angular momentum will be concentrated near the inner edge of the disc, making the estimates above invalid. This problem is discussed in more detail in Appendix \ref{A:disc_size}. We plan to study how a realistic cooling model (together with heating by gravitational instability) sets the thermal profile of an accreting, gravitationally unstable disc in future work.

\section{Numerical convergence}
\label{sec:convergence}
\begin{figure}
    \centering
    \includegraphics[width=.4\textwidth]{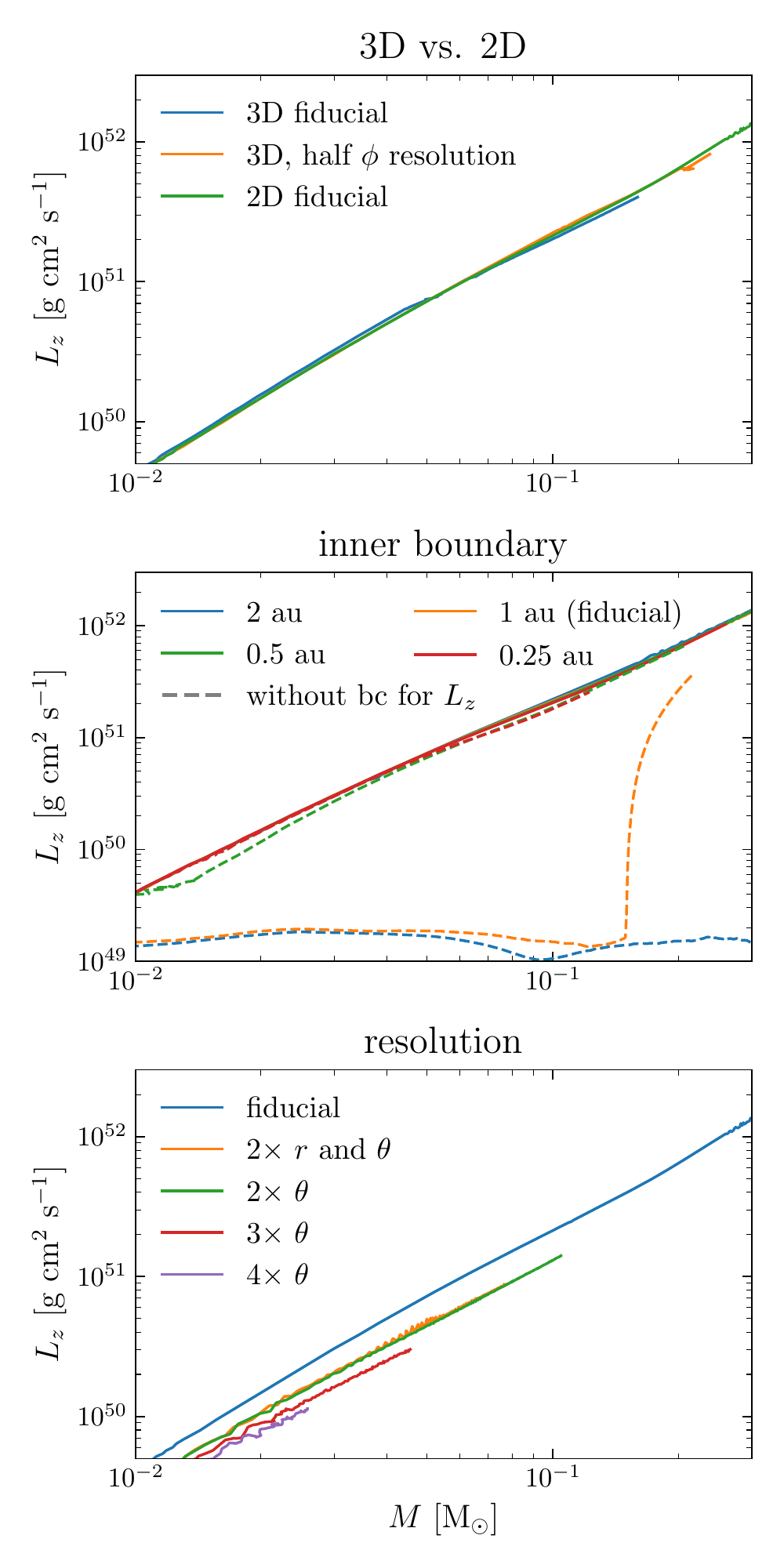}
    \caption{Mass--angular-momentum relation of the protostar-disc system for different numerical setups. Top panel: comparison between 3D simulations with different $\phi$ resolutions and a 2D simulation. Middle panel: 2D simulations with different inner boundary sizes. Simulations without the $L_z$ flux boundary condition are also plotted as dashed lines. Bottom panel: different resolutions in $r$ and $\theta$. (See more discussion on $\theta$ resolution in text.) We also test the dependence on the value of numerical caps of diffusivity and velocity, but the results are not shown here since there is no visible difference.}
    \label{fig:ML_convergence}
\end{figure}

\begin{figure}
    \centering
    \includegraphics[width=.5\textwidth]{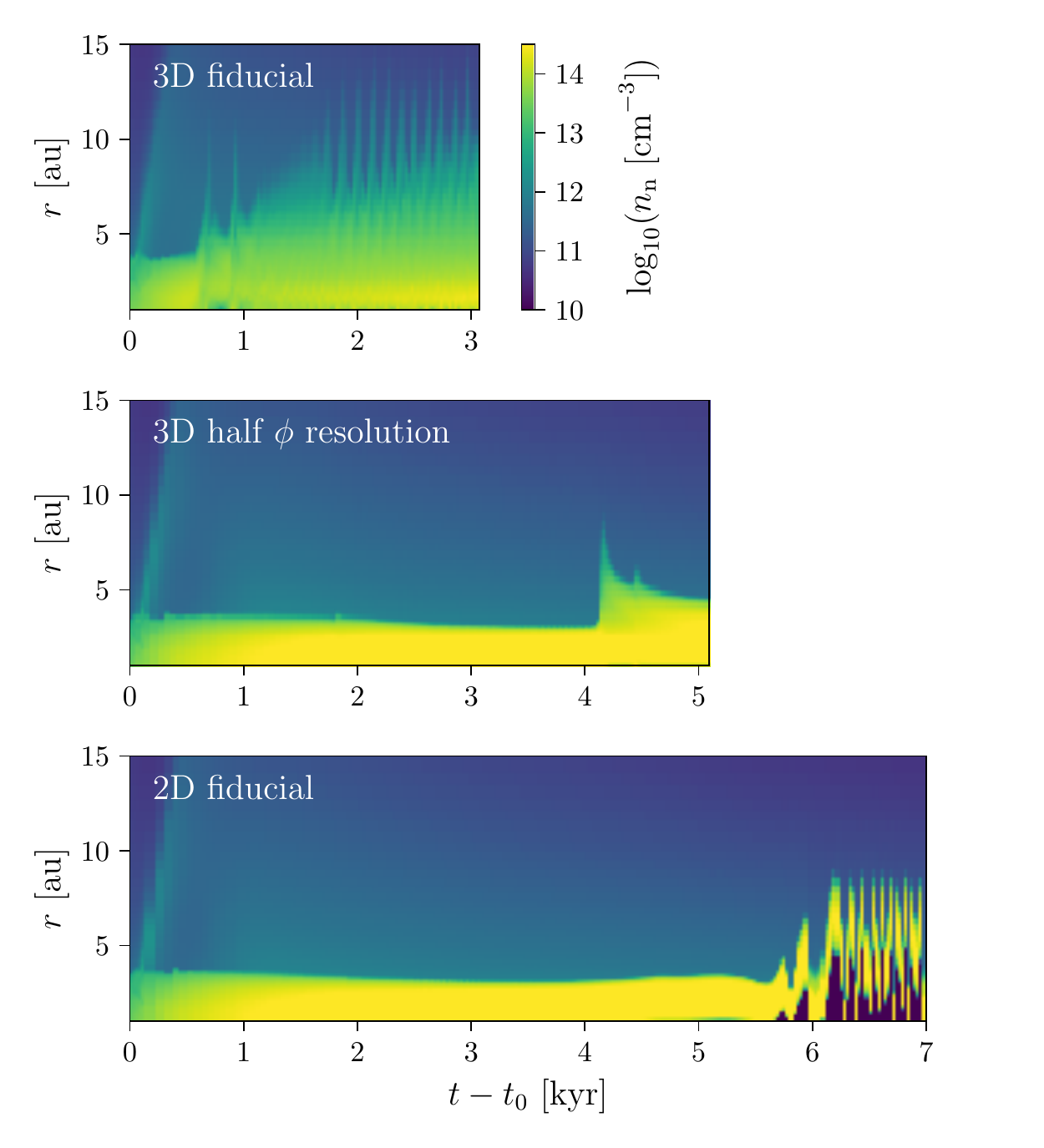}
    \vspace{-1em}
    \caption{Space-time plots of midplane density for simulations with different $\phi$ resolution. To align the time axes, we choose a reference time $t_0$ defined as the epoch when $M_{10}=0.01M_\odot$. Lowering $\phi$ resolution delays the appearance of gravitationally excited spirals, and for an  axisymmetric simulation gravitational instability leads to radial disc oscillations instead of spreading. Directly resolving gravitational instability requires a $\phi$ resolution no lower than our fiducial value.}
    \label{fig:convergence_nt}
\end{figure}

In previous sections we focused on analyzing results from the fiducial 3D simulation. Here we compare this fiducial simulation to a set of 3D and 2D simulations to discuss whether our results are sensitive to numerical parameters such as resolution and inner boundary size. We find that details of the 3D disc evolution may still be sensitive to resolution, but the $M$--$L_z$ relation of the protostar-disc system is robust against changing numerical parameters. We also illustrate that convergence on the $M$--$L_z$ relation can only be achieved if the inner boundary is very small or an angular-momentum flux boundary condition similar to ours is applied.

\subsection{2D vs. 3D and $\phi$ resolution}
\label{subsec:phi_res}

First we compare our fiducial 3D simulation with a 3D simulation at half $\phi$ resolution and an axisymmetric 2D simulation. All three simulations have nearly identical setup, except the $\phi$ resolution.
In the top panel of Fig.~\ref{fig:ML_convergence} we compare their $M$--$L_z$ relations, and find them to be nearly identical. This good agreement is not surprising, since in Section \ref{sec:ML} we concluded that the mass and angular-momentum budget of the protostar-disc system is mainly controlled by accretion through the pseudo-disc, which is axisymmetric and therefore insensitive to $\phi$ resolution.

The details of disc evolution, however, can look very different between different $\phi$ resolutions and between 3D and 2D, as shown in Fig.~\ref{fig:convergence_nt}.
In both 3D simulations, we see disc spreading by gravitational instability. But for low $\phi$ resolution, the onset of gravitational instability (marked by the sudden increase in disc size) occurs much later and the disc size is smaller. Disc spreading is also more bursty in this case: the disc suddenly increases size in an episode of strong gravitational instability, then has to wait a relatively long time before it becomes unstable again (which is not covered by our simulation).
These differences are mainly because the low $\phi$ resolution suppresses non-axisymmetric perturbations and a disc has to reach smaller $Q$ to excite spiral waves and initiate disc spreading.
For 2D, the disc no longer spreads but undergoes radial oscillations, because the gravitationally excited spiral waves that transport angular momentum are no longer allowed in axisymmetry.
In summary, correctly capturing disc spreading by gravitational self-regulation through direct simulation requires a $\phi$ resolution no lower than our fiducial value.

As a side note, certain details of disc evolution, such as the amplitude of spiral waves and the exact $Q$ value, could require a much higher resolution to fully converge. Local shearing-box simulations find that gravitational instability creates small-scale perturbations (`gravitoturbulence'), which can affect (and sometimes disrupt) the large-scale spiral waves \citep{Gammie2001,Riols2017}. The amplitude of spiral waves and the relation between $Q$ and the rate of angular-momentum transport requires at least 8--16 cells per scale height to fully converge; such resolution is basically impossible for 3D disc-formation simulations in the near future.
Still, these details should not affect the basic picture that the disc spreads at marginal gravitational instability.

There is another interesting conclusion one can draw from this comparison. The similarity in the $M$--$L_z$ relation even when disc evolution is very different in 2D and 3D implies a lack of feedback. In other words, the protostar-disc system cannot affect the pseudo-disc accretion process, which determines the $M$--$L_z$ relation.

\begin{figure}
    \centering
    \includegraphics[width=.5\textwidth]{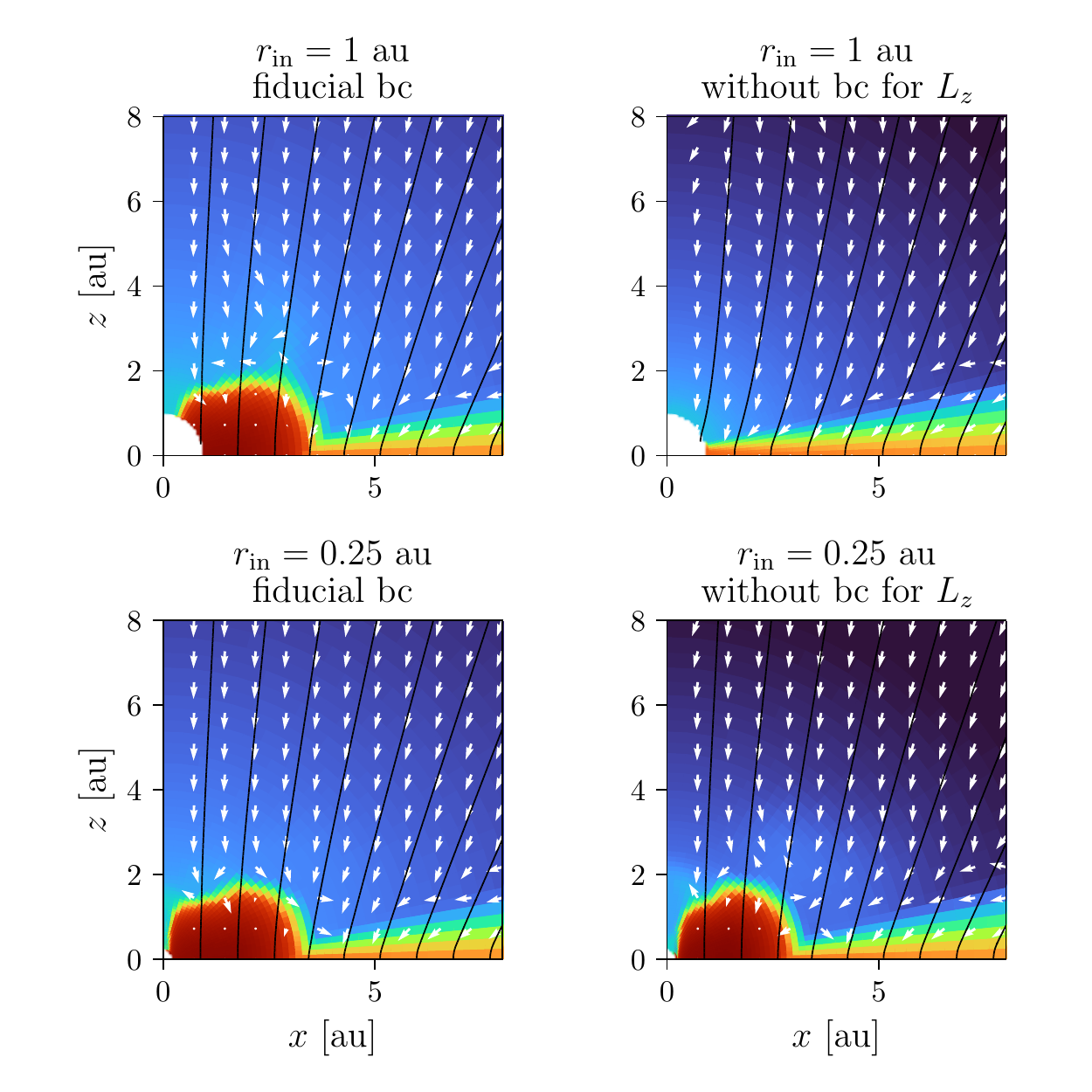}
    \vspace{-2em}
    \caption{Snapshots at $M_{10}=0.1M_\odot$ for 2D simulations with different $r_{\rm in}$ and inner boundary condition. The plotting scheme and colour scale are the same as in Fig.~\ref{fig:snapshot_overview}. Simulations show good agreement if $r_{\rm in}$ is sufficiently small or the boundary condition on $L_z$ flux is applied. When $r_{\rm in}$ is relatively large and the $L_z$ flux boundary condition is not used (top right panel), disc formation is fully suppressed.}
    \label{fig:convergence_torus}
\end{figure}

\subsection{Inner boundary size and boundary condition}

Now we consider the effect of the inner boundary size $r_{\rm in}$ and inner boundary condition. In Fig.~\ref{fig:ML_convergence} we show the $M$--$L_z$ relation of a set of 2D simulations with $r_{\rm in}$ ranging from $0.25$ au to 2 au. We see a very good agreement, even when the largest $r_{\rm in}$ we use is already comparable to the size of the first hydrostatic core/torus.

To test whether our inner boundary condition is correct and necessary, we run another set of 2D simulations with a different boundary condition where we do not force the angular-momentum flux through the inner boundary to be zero. The result for the two different boundary conditions are similar for $r_{\rm in}\lesssim 0.5~{\rm au}$, suggesting that our boundary condition does not introduce any artifact that affects the angular-momentum budget of the system.
For larger $r_{\rm in}$, however, simulations without the angular-momentum flux boundary condition show very poor numerical convergence and much lower $L_z$. And for $r_{\rm in}\geq 1$ au, removing the angular-momentum flux boundary condition makes the torus disappear completely (Fig.~\ref{fig:convergence_torus}). Therefore, having a boundary condition that limits the angular-momentum flux through the inner boundary is necessary, unless one can afford an inner boundary size as small as $0.5~{\rm au}$ or less.

The necessity of having a small inner boundary or limiting the angular-momentum flux through the inner boundary has a physical origin. The mean specific angular momentum coming from the pseudo-disc is quite low; for our fiducial simulation, the circularization radius is ${\sim}1~{\rm au}$. (Note that $L_z\propto M^{1.5}$ gives approximately constant circularization radius.) In order to form a disc, mass has to be accumulated around this circularization radius until the torus becomes gravitationally unstable and starts transporting angular momentum. However, when the inner boundary is larger than this circularization radius and angular momentum can be freely advected through the inner boundary, all infalling angular momentum will be lost through the inner boundary and the disc cannot form.

{Additionally, the two boundary conditions compared here correspond to two limiting scenarios of angular-momentum transport at $r<r_{\rm in}$. Our fiducial boundary condition with zero $L_z$ flux effectively assumes that all angular momentum that gets within $r_{\rm in}$ eventually returns to $r>r_{\rm in}$; removing the $L_z$ flux boundary condition effectively assumes that all angular momentum that gets past $r_{\rm in}$ is eventually removed from the protostar-disc system. The fact that these two boundary conditions give very similar evolution when $r_{\rm in}\lesssim 0.5~{\rm au}$ suggests that the evolution is not sensitive to the details of the angular-momentum transport at small radii (${\lesssim}0.5~{\rm au}$). One caveat is that these two boundary conditions both implicitly assume that nothing occurring inside of the inner boundary can affect angular-momentum transport in the active domain.}

\subsection{Resolution in $r,\theta$}

Finally, we test convergence with respect to $r,\theta$ resolution using a set of 2D simulations with different resolutions. The resulting $M$--$L_z$ relations are plotted in the bottom panel of Fig.~\ref{fig:ML_convergence}. The convergence in $r$ resolution is good, as suggested by the comparison between the green and orange curves.
The result also converges as $\theta$ resolution increases, but the convergence is slower. We find that this is associated with an numerical artifact in the polar region. We discuss this artifact in Appendix \ref{A:polar_EMF} and introduce a simple correction for 2D simulations that can remove this artifact. Comparing results with and without this correction in Appendix \ref{A:polar_EMF}, we find that the $4\times$ $\theta$ resolution run gives a good estimate of the actual $M$--$L_z$ relation, which has $L_z$ about a factor of 2 smaller than the fiducial resolution runs.

In summary, after inspecting convergence with respect to all numerical parameters (including resolution), we are confident that the $M$--$L_z$ scaling in our fiducial simulation (Fig.~\ref{fig:ML}) has the correct shape and is accurate to within a factor of 2.

\section{Summary and discussion}\label{sec:summary}

In this paper we use non-ideal MHD simulations to investigate the evolution of a pre-stellar core until a few kyr after protostar formation. We observe the formation of an initially small disc, which then spreads by gravitational instability to ${\sim}30~{\rm au}$ in diameter (\S \ref{sec:evolution}). Here we summarize the main results of our simulations in terms of the physical picture of disc formation and requirements for numerical convergence, and discuss how our results connect to recent observational estimates of protostellar disc masses. The robustness of our results with respect to physical initial conditions and disc chemistry will be evaluated in a parameter study in a subsequent publication. 

\subsection{Disc formation: a simple physical picture}

Broadly speaking, the formation and evolution of the protostellar disc is determined by two processes: the injection of mass and angular momentum to (and the removal from) the protostar-disc system, and the redistribution of angular momentum within the protostar-disc system.

Regarding the first process, we find that the mass and angular momentum of the protostar-disc system is determined mainly by injection from a pseudo-disc in near free-fall, whose specific angular momentum is low due to magnetic braking (\S \ref{sec:ML}).
The removal of mass and angular momentum by outflow and magnetic braking in the protostar-disc system is negligible (contrary to some earlier studies that may have underestimated magnetic diffusivity in disc),
and there is barely any feedback from the protostar-disc system that affects the pseudo-disc injection rate (\S \ref{subsec:budget}, \S \ref{subsec:phi_res}).
In other words, disc formation is hierarchical: the mass and angular-momentum budget of the small scale (protostar-disc system) is solely determined by large scale (collapse of pre-setllar core and pseudo-disc evolution).

Regarding the second process, we find that the redistribution of angular momentum within the protostar-disc system is mainly facilitated by gravitational instability and the associated non-axisymmetric angular-momentum transport (\S \ref{subsec:transport}).
Transport by gravitational instability serves as a self-regulation mechanism that holds most of the disc marginally unstable, with a Toomre $Q \sim 1$--$2$ (\S \ref{subsec:self_regulation}). For a given mass and angular momentum of the protostar-disc system, the size and column density profile of the disc is determined by this gravitational self-regulation and can be estimated analytically if the thermal profile of the disc is known or can be constrained (\S \ref{subsec:disc_size}). Another important implication is that the specific angular momentum of the disc can be much larger than of the material being accreted by the disc, allowing the formation of large discs even when magnetic braking in the pseudo-disc is strong.

Together, we now have a relatively simple picture of disc formation. Large-scale processes (collapse of the pre-stellar core and pseudo-disc evolution), which are mostly laminar and axisymmetric, determine the evolution of total mass and angular momentum of the protostar-disc system. The total mass and angular momentum of the protostar-disc system then directly determine the properties of the disc, including size and density profile, through gravitational self-regulation.

{Of course, this physical picture is based on simulations that cover only the first several kyr of disc evolution, and which use one particular set of initial conditions and chemical abundances.}
In future studies, we will determine if this picture is still applicable for the long-term evolution of discs throughout the Class 0/I phase and for different initial conditions {and dust populations}.

\subsection{Numerical convergence: the importance of the inner boundary}

Poor numerical convergence has been a significant problem for simulations of protostellar disc formation with relatively low resolution. We use a set of 2D and 3D simulations to study the numerical convergence of our results for all numerical parameters (including resolution) in our model, and find relatively good convergence for the mass and angular-momentum budget of the protostar-disc system.

Our relatively good numerical convergence is mainly due to two reasons. First, we use a sufficiently small inner boundary $r_{\rm in}$, which helps to conserve angular momentum and resolve the early evolution of the disc. We use $r_{\rm in}$ as small as 0.25~au for 2D simulations and 1~au for 3D simulations, which is significantly smaller than used in most previous studies.\footnote{In simulations using a Cartesian grid, our small $r_{\rm in}$ corresponds to both a small sink particle and sufficiently high resolution close to the protostar.} Second, we use an inner boundary condition that limits the angular-momentum flux through the inner boundary, which relaxes the requirement on $r_{\rm in}$ for numerical convergence by at least a factor of a few (which translates to more than an order of magnitude in computational cost; see more discussion on computational cost in Appendix \ref{A:cost}).

We also explain the origin of this sensitive dependence on the inner boundary: disc formation requires angular momentum to be accumulated first near the protostar and then be transported outwards by gravitational instability. If the inner boundary is too large and angular momentum can flow freely through the inner boundary, the initial accumulation of angular momentum is prohibited and disc formation is strongly suppressed.\footnote{For a Cartesian grid, one must also ensure that angular momentum is not appreciably lost to the grid when the angular sizes of the cells are large. This is especially problematic when a significant amount of angular momentum may have to remain near the protostar for many orbits before being transported outwards.}

\subsection{Observational hints of gravitational self-regulation}
\label{subsec:summary_obs}

Our simulation suggests that protostellar discs are likely marginally gravitationally unstable for at least a significant fraction of Class 0/I phase. Here we discuss how this prediction compares with recent observations of young discs.

The most direct way of determining whether a disc is gravitationally unstable is through estimating its Toomre $Q$.
This requires estimating disc mass, which for Class 0/I discs is usually inferred from dust continuum emission.
However, such a measurement typically assumes optically thin emission at the observed wavelength, which is likely untrue for some systems \citep{GalvanMadrid2018, Liu2020} and may lead to significant underestimation of disc mass.
To calibrate this effect, one has to observe the disc at a very long wavelength to ensure optically thin emission, or multiple wavelengths to constrain opacity.
For example, \citet{Sharma2020} use ALMA 0.87-mm data and VLA 8-mm data to show that the disc around an outbursting Class 0 protostar likely has $Q<1$ before outburst. They also comment that using the ALMA data (more optically thick because of the shorter wavelength) alone would underestimate disc mass by at least a factor of 10.
A similar trend is also visible in the VANDAM survey of Orion protostars \citep{Tobin2020}, where ALMA 0.87-mm data suggest most discs have $Q$ well above unity (assuming optically thin dust emission), but VLA 8-mm data, when available, gives much smaller $Q$ values that are usually consistent with marginally unstable discs (see their Table 9).
It is also worth noting that the protostar mass in these systems are usually unknown. The estimates discussed above both assume a certain fiducial protostar mass when calculating $Q$, and that leads to large uncertainty. (On the other hand, if gravitational self regulation is indeed applicable to a wide range of Class 0/I systems, one can use disc mass observations to put rough constraints on protostar mass.)
Overall, although there are several important uncertainties, observed Class 0/I disc masses, when the dust optical depth is properly accounted for, should be in broad agreement with our prediction.

Less direct evidence for gravitationally unstable circum-protostellar discs in the Class 0/I stage may be obtained via measurements of Toomre $Q$ in Class II discs. If discs are marginally unstable in the Class 0/I phase, then it is reasonable to expect young Class II discs often to have $Q$ of a few. Early measurements of Class II discs usually use CO (or CO isotope) lines to estimate gas mass, or use dust continuum to estimate dust mass. These studies \citep[e.g.,][]{Ansdell2016} generally find that Class II discs are not very massive, with disc-to-star mass ratios of a few percent or less.
However, these estimates suffer from uncertainties such as CO/H$_2$ ratio, dust-to-gas ratio, and optical depth of observed lines or wavelength, and such uncertainties often lead to systematic underestimation of disc mass.
A recent study by \citet{Powell2019} discusses these issues and proposes a new method for constraining disc mass using dust disc visibility at multiple wavelengths. This method is immune to the aforementioned uncertainties and tends to give much higher disc mass: Out of the 7 discs modeled in this study, 6 show $Q\lesssim 3$.
Observations using $^{13}$C$^{17}$O, a rare isotope, also suggest that earlier studies may have significantly underestimated disc opacity and mass, and the two discs observed with this method are both likely gravitationally unstable \citep{Booth2019,Booth2020}.
Therefore, it is likely that typical young Class II discs have once been (or still are) gravitationally unstable.

\section*{Acknowledgments}

It is a pleasure to thank Patrick Hennebelle, Jim Stone, and Kengo Tomida for useful discussions; the referee for constructive feedback; and especially Shantanu Basu for comments on a draft version of this manuscript. The simulations presented in this article were performed on computational resources managed and supported by Princeton Research Computing, a consortium of groups including the Princeton Institute for Computational Science and Engineering (PICSciE) and the Office of Information Technology's High Performance Computing Center and Visualization Laboratory at Princeton University. Partial support was provided by an Alfred P.~Sloan Research Fellowship in Physics to M.W.K.
\vspace{-1ex}

\section*{Data Availability}
The data underlying this article will be shared on reasonable request to the authors.

\bibliographystyle{mnras}
\bibliography{XK20}

%\newpage
%~
%\newpage

\appendix

\section{Details of numerical setup}
\label{A:numerical}
\subsection{Self gravity}
The gravitational potential in the domain can be separated into two components, one corresponding to the mass in the domain, and the other the point accretor $M_{\rm acc}$:
\begin{equation}
\Phi = \Phi_{\rm s} + \Phi_{\rm acc}.
\end{equation}
Here $\Phi_{\rm acc} = -GM_{\rm acc}/r$ and $\Phi_{\rm s}$ is obtained by solving the Poisson equation
\begin{equation}
\nabla^2\Phi_{\rm s} = 4\pi G(\rho-\rho_\infty).
\label{eq:poisson}
\end{equation}
We adopt an open boundary condition for equation \eqref{eq:poisson}: $\Phi_{\rm s}$ is finite at $r\to 0$ and approaches zero at $r\to\infty$ if we were to continue solving $\nabla^2\Phi=0$ outside the domain.

On the right-hand side of equation \eqref{eq:poisson}, we have replaced the usual density term $\rho$ with $(\rho-\rho_\infty)$, where $\rho_\infty$ is the background mass density corresponding to the background number density $n_\infty$ in equation \eqref{eq:n}. We also replace $\rho\boldsymbol{g}$ with $(\rho-\rho_\infty)\boldsymbol{g}$ in the momentum equation. This modification, together with an open boundary condition for $\Phi_{\rm s}$,  corresponds physically to a situation where the system is embedded in an infinite uniform ambient density $\rho_\infty$. In reality, protostars are not in vacuum but are embedded in dense molecular gas (with density ${\sim}\rho_\infty$), and so this `Jeans swindle' modification is likely a better approximation of reality than the standard Poisson equation.

We solve equation \eqref{eq:poisson} numerically in spherical-polar coordinates using spherical harmonic decomposition. The transformations to and from the spherical harmonic basis are performed using a fast Fourier transform (FFT) in the $\phi$ direction (necessary only for 3D simulations) and an inner product with the associated Legendre polynomials through direct summation. In this basis of spherical harmonics, the Poisson equation becomes a second-order ordinary differential equation for each $Y^{lm}$ component. The (second-order accurate) finite-difference version of this differential equation corresponds to a tri-diagonal matrix equation, which we solve using LU factorization with pre-computed coefficients. This allows a solution cost of $\mathcal O(N_r)$ for each $Y^{lm}$ component.
Overall, the computational cost for a single solve is $\mathcal O(N) + \mathcal O(N\log N_\phi) + \mathcal O(N N_\theta)$, where $N_r$, $N_\theta$, $N_\phi$ are number of cells in each direction and $N \equiv N_r N_\theta N_\phi$ is the total number of cells. On paper, this scaling does not look very good. In practice, however, the simplicity of the algorithm lends it good performance, and the cost of solving $\Phi_{\rm s}$ becomes significant (compared to everything else) only when $N_\theta\gtrsim 100$. At our fiducial resolution, the cost of the self-gravity solver constitutes only ${\sim}15\%$ of the total computational cost.\footnote{As a side note, $\mathcal O(N)$ algorithms such as multigrid tend to become much less efficient when there are too few cells per thread (e.g., when each thread owns $8^3$ or $16^3$ cells); this problem is less severe for our spherical harmonic solver. This makes our solver suitable for reducing the real-world time of a simulation by using more cores.}

\begin{figure}
    \centering
    \vspace{-2em}
    \includegraphics[width=.5\textwidth]{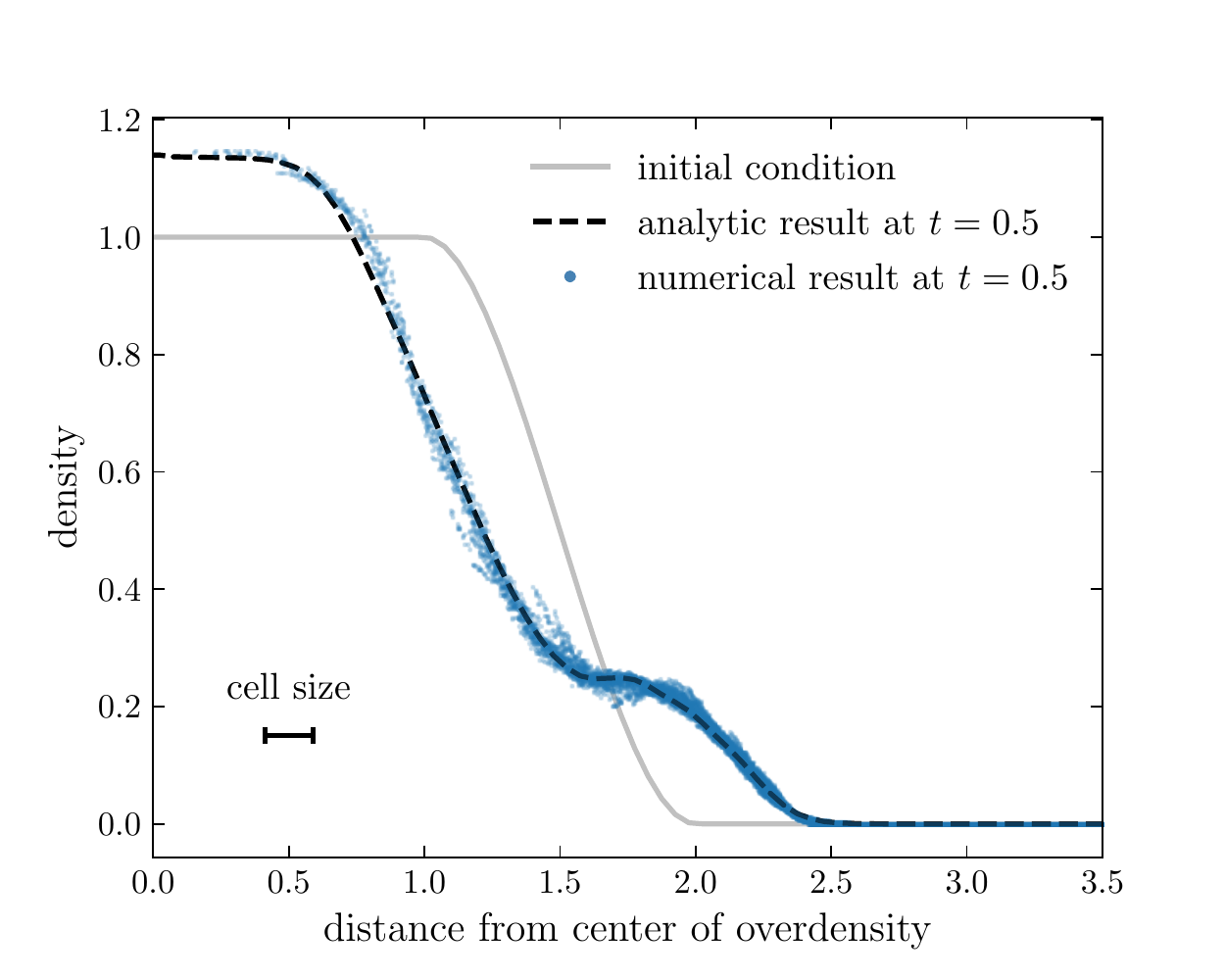}
    \vspace{-1em}
    \caption{Test problem for our self gravity algorithm. We simulate the collapse of a spherical overdensity whose centre is not at the coordinate origin. The numerical result shows good agreement with analytic prediction. The initial condition and typical cell size in the overdense region are also shown for reference.}
    \label{fig:test_time_evolution}
\end{figure}

To test our implementation of self gravity, we simulate the collapse of an isothermal spherical overdensity whose centre is not at the coordinate origin. We set $4\pi G=1$ and isothermal sound speed $c_s=1$.
The overdensity is centred at $(r,\theta,\phi) = (4,1,0.7)$, with initial density profile given in Figure \ref{fig:test_time_evolution}.
At a resolution of ${\rm d}r/r\approx {\rm d}\theta\approx 0.05$ (comparable to our fiducial resolution) and ${\rm d}\phi\approx 0.1$, the evolution agrees well with analytic results, and the overdensity remains spherical during the collapse without visible artifacts (Figure \ref{fig:test_time_evolution}).

\subsection{Velocity and diffusivity caps}

To avoid very short timesteps, we apply caps on the magnetic diffusivities (both Ohmic and ambipolar), the flow speed $v$, and the Alfv\'en speed $v_{\rm A}$. The cap on $\eta$ is  applied directly when calculating $\eta$ at each timestep, and the caps on $v$ and $v_{\rm A}$ are applied by increasing the density when the velocity magnitude is greater than the cap value (while holding momentum and magnetic field constant). In other words, the velocity cap is implemented as a density floor.

These caps are often applied to star-formation simulations, but sometimes their values are chosen somewhat arbitrarily. Here we try to give a more physical choice of cap values that in principle should not affect the dynamics of the protostar-disc system. In order to do this, we choose the caps so that the characteristic timescale associated with the diffusivity cap, $r^2/\eta_{\rm cap}$, is at most comparable to the free-fall timescale; and the characteristic timescale associated with the velocity cap, $r/v_{\rm cap}$, is much smaller than the free-fall timescale. (Note that both free-fall timescale and Keperian timescale are $\sim\sqrt{-g_r/r}$.) The rationale for the first criteria is that, when the diffusion timescale is smaller than the gravitational (free-fall, or Keplerian) timescale, the field is largely decoupled from the gas and the behaviour is not sensitive to the exact value of $\eta$.
We also want to choose the cap values such that the CFL timestep corresponding to the caps are similar in the whole domain. This allows us to choose higher caps for larger $r$ without affecting the timestep. Since the cell size are $\propto r$, we choose $\eta_{\rm cap}\propto r^2$ and $v_{\rm cap}\propto r$.
Combining the two motivations above, we choose the following caps for diffusivity and velocity:
\begin{equation}
\eta_{\rm cap} = f_\eta\tau_g^{-1}r^2,~~~
v_{\rm cap} = f_v\tau_g^{-1}r.
\end{equation}
Here $\tau_g^{-1}$ is defined as the global maximum of $\sqrt{-g_r/r}$ on the midplane, and $f_\eta,f_v$ are constant factors. Our default choices are $f_\eta=1$ and $f_v=50$. To check that these caps do not affect the dynamics, we re-ran some 2D simulations with $f_v=100$ and with $f_\eta = 0.5$, $2$, and $4$. All of these runs show nearly identical behaviour.

\subsection{Polar averaging}

One major disadvantage of using spherical-polar coordinates for 3D simulations is that the cells become narrow wedges towards the pole, which may limit the timestep severely. This problem is usually tackled using mesh refinement near the midplane, which allows the $\phi$ resolution near the midplane to be higher than that around the pole. However, this choice is not suitable for our simulation since our self gravity solver is not (yet) compatible with mesh refinement.

To circumvent this problem, we need to `de-refine' cells near the pole in the $\phi$ direction. This is achieved by performing a `polar averaging'. That is, after each update of cell quantities we perform a Fourier transform in the $\phi$ direction for cells near the pole and truncate the high-frequency terms of density, momentum, and magnetic field. (For magnetic field, we do this to $B_r$ and $B_\theta$, and then $B_\phi$ follows from $\grad\bcdot\bb{B}=0$.) We also perform the same truncation for the EMF before using it to update the magnetic field. The number of terms kept, $n_\phi$, is chosen such that the effective cell size in the $\phi$ direction,
\begin{equation}
\rmd\phi^{\rm eff} = r\sin\theta (\rmd\phi) N_\phi/n_\phi,
\end{equation}
is no smaller than $r(\rmd\theta)_{\rm min}$. Here $\theta$ is the poloidal angle of the cell centre, and $\rmd\theta$ and $\rmd\phi$ are the mesh spacings in the poloidal and azimuthal directions; $(\rmd\theta)_{\rm min}$ is the minimum $\rmd\theta$, which is located in the midplane. We then use this effective cell size to replace the actual cell size when computing the CFL condition for the next timestep.
This averaging procedure is necessary only when we need $n_\phi< N_\phi$ to get $\rmd\phi^{\rm eff}\geq r(\rmd\theta)_{\rm min}$.
For example, in our fiducial 3D simulation, the averaging is only applied to the first two cells in the $\theta$ direction near the pole.

\section{Diffusivity for our dust profile}
\label{A:diffisivity}

\begin{figure}
    \centering
    \hbox{\hspace{-1em}
    \includegraphics[width=.5\textwidth]{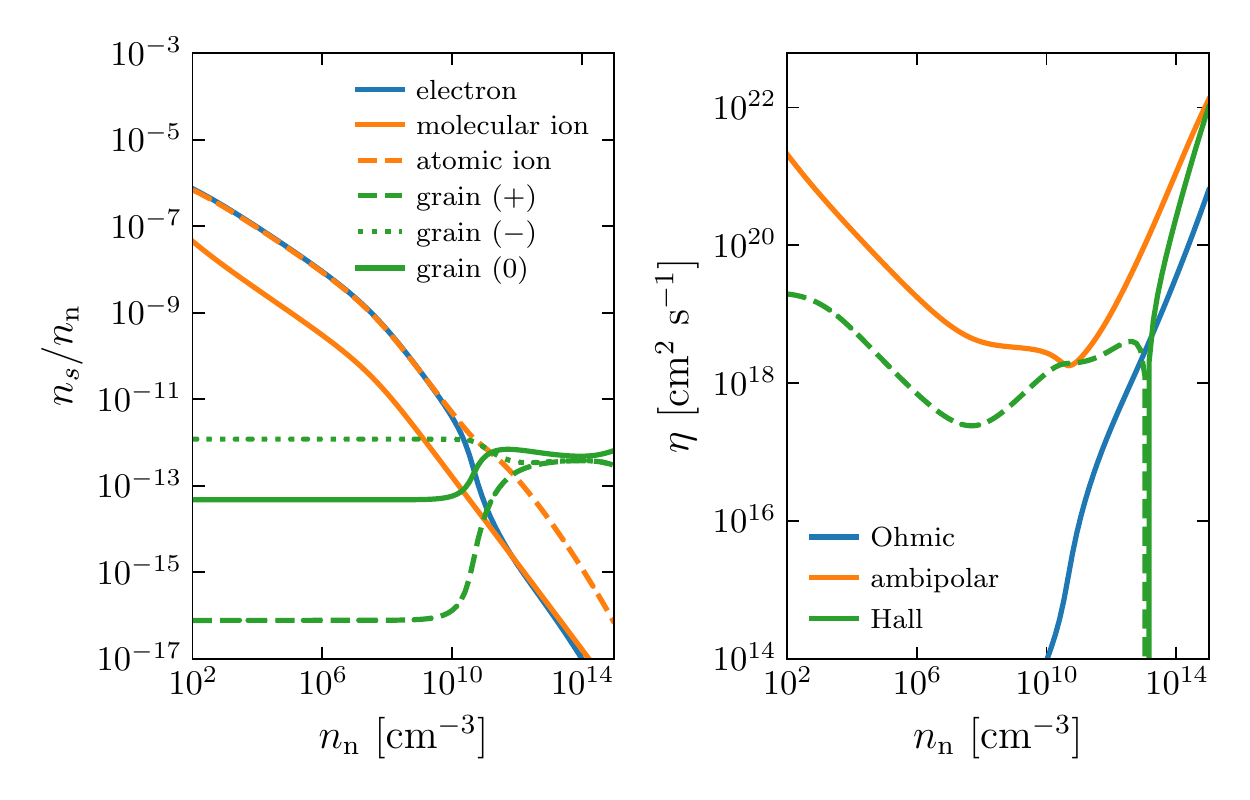}}
    \vspace{-1em}
    \caption{Left: Fractional abundances of species for our chemical network. Right: Magnetic diffusivities as a function of density, assuming $B = 0.1~\mu{\rm G}~ (n_{\rm n}/{\rm cm}^{-3})^{1/2}$. (The Hall effect is not included in our simulations.)}
    \label{fig:abundance_and_eta}
\end{figure}

\begin{figure}
    \centering
    \includegraphics[width=.5\textwidth]{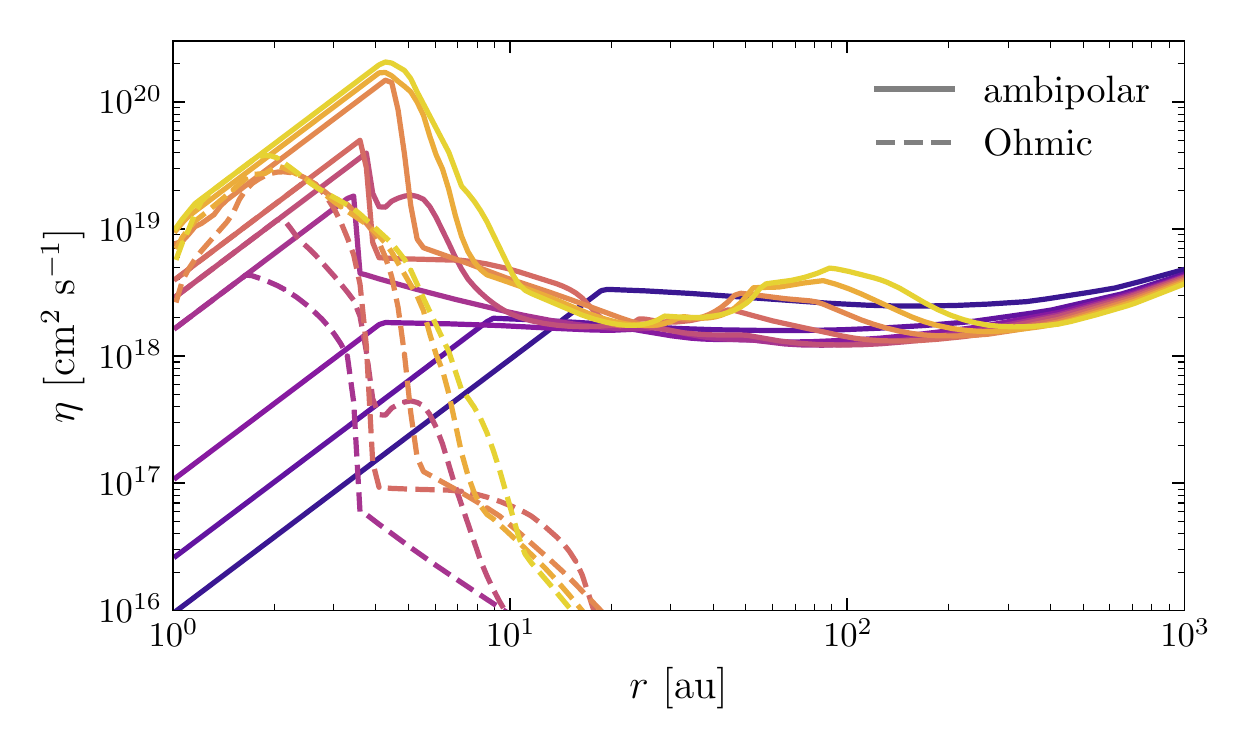}
    \vspace{-1em}
    \caption{Midplane magnetic diffisivity $\eta$ of the fiducial 3D simulation at different epochs. The epochs plotted and the colour scheme are the same as in Fig.~\ref{fig:curves_overview}. 
    The numerical cap on $\eta$, which is only applied when the diffusion timescale is shorter than dynamical timescale, is evident at small $r$.}
    \label{fig:Rm}
\end{figure}

In Fig.~\ref{fig:abundance_and_eta} we show the fractional abundances of the species in our chemical network and the resulting magnetic diffusivities as functions of the neutral number density. Because magnetic diffusivities in general depend also on the strength of the magnetic field, for the purposes of this figure we take $B = (n_{\rm n}/100~{\rm cm}^{-3})^{1/2}~\mu{\rm G}$, which is similar to the midplane $B$-$n_{\rm n}$ relation during the pre-stellar collapse phase of our simulations. We also plot the midplane magnetic diffusivities in our fiducial 3D simulation in Fig.~\ref{fig:Rm}.

For the range of density in which we are interested here, ambipolar diffusion is the dominant non-ideal effect. Also note that our choice of dust size gives significantly higher diffusion at intermediate densities (${\sim}10^7$--$10^{10}~{\rm cm}^{-3}$) compared to the standard MRN profile \citep[cf.][]{Zhao2018diffusivity}.

\section{Estimating the size and mass of a non-isothermal disc}
\label{A:disc_size}

In Section \ref{subsec:disc_size} we estimated the size and mass of the disc under the assumption that disc is isothermal and marginally gravitationally unstable. However, the resulting disc surface density from equation \eqref{eq:Sigma_est} is ${\propto}R^{-3/2}$, suggesting that the isothermal assumption may no longer be good at small cylindrical radius $R$ where disc column density is high. In this appendix we discuss the effect of a non-isothermal inner disc.

For a non-isothermal disc, we can still apply equation \eqref{eq:Sigma_est}, except now $c_{\rm s0}$ must be replaced by an $R$-dependent sound speed $c_{\rm s}(R)$. Suppose $c_{\rm s}(R)\propto R^{-\beta}$ at small $R$. A marginally gravitationally unstable Keplerian disc would satisfy $\Sigma\propto R^{-\frac 32-\beta}$ and
\begin{align}
&\frac{\rmd M_{\rm d}}{\rmd\ln R} \propto R^{\frac 12-\beta},\\
&\frac{\rmd L_{\rm d}}{\rmd\ln R} \propto R^{1-\beta}.
\end{align}
For sufficiently large $\beta$ ($\beta\geq 1/2$ for mass and $\geq 1$ for angular momentum), the disc mass and angular momentum diverge if we integrate the above equations to $R\to 0$. Physically, this means that most of the mass and angular momentum are concentrated near the inner edge of the marginally unstable region ($R_{\rm d,in}$), and that the total disc mass and size are very sensitive to the exact value of $R_{\rm d,in}$. For example, when $\beta=1$, $\rmd L_{\rm d}/\rmd \ln R\sim{\rm const}$ and $R_{\rm d}\propto R_{\rm d,in}$ for a given $M$ and $L_z$.

Now we estimate the value of $\beta$ for the EoS used in our simulation. Since we assume a geometrically thin disc with $Q=Q_0\sim 1$ (i.e., self gravity and pressure support are comparable), the disc scale height is ${\sim}c_{\rm s}/\Omega$ and $\Sigma\sim \rho c_{\rm s}/\Omega$, where $\rho$ is the midplane mass density. Using equation \eqref{eq:Sigma_est}, we find
\begin{equation}
\rho \sim \frac{GM}{\pi G Q_0 R^3}.
\end{equation}
For the EoS used in our simulations, at high density $p\propto \rho^{\gamma}$ with $\gamma=5/3$, so that $c_{\rm s}\propto \rho^{1/3} \propto R^{-1}$ and $\beta=1$. In this case, both the disc mass and size depend sensitively on $R_{\rm d,in}$. In our 3D simulation $R_{\rm d, in}$ is similar to, and probably controlled by, the size of the inner boundary of the computational domain; it is therefore possible that the disc size will be different for a different inner boundary size for the particular barotropic EoS we adopt.


We can also ask whether one should expect such a large $\beta$ in reality. The $\gamma=5/3$ EoS at high density corresponds physically to a trivial level of cooling. In reality, however, the cooling rate may still be nontrivial in an optically thick, gravitationally unstable disc. This is in part because the mixing due to the turbulent motion from gravitational instability gives very effective convection (mixing time is ${\lesssim}\Omega^{-1}$), and for midplane temperature $T$ the cooling per unit area will be ${\sim}\sigma T^4$, as opposed to ${\sim}\sigma T^4/\tau$ for an optically thick, non-convective disc with optical depth $\tau$. When a nontrivial amount of cooling is present, the EoS will be less steep; and $\gamma$ just needs to be slightly less steep than $5/3$ to have $\beta>1$ and a disc size insensitive to $R_{\rm d,in}$.
Meanwhile, although most of the disc mass will still be near the inner edge when $\gamma$ is close to 5/3, the density profile in the outer part of the disc remains insensitive to the location of the inner edge.
Therefore, when disc evolution is regulated by gravitational instability and a realistic thermal profile is assumed, our estimate for disc size and density profile should be insensitive to the exact location of the disc inner edge.

\section{Polar correction}
\label{A:polar_EMF}

\subsection{Problem description and possible origin}

\begin{figure}
    \centering
    \includegraphics[width=.5\textwidth]{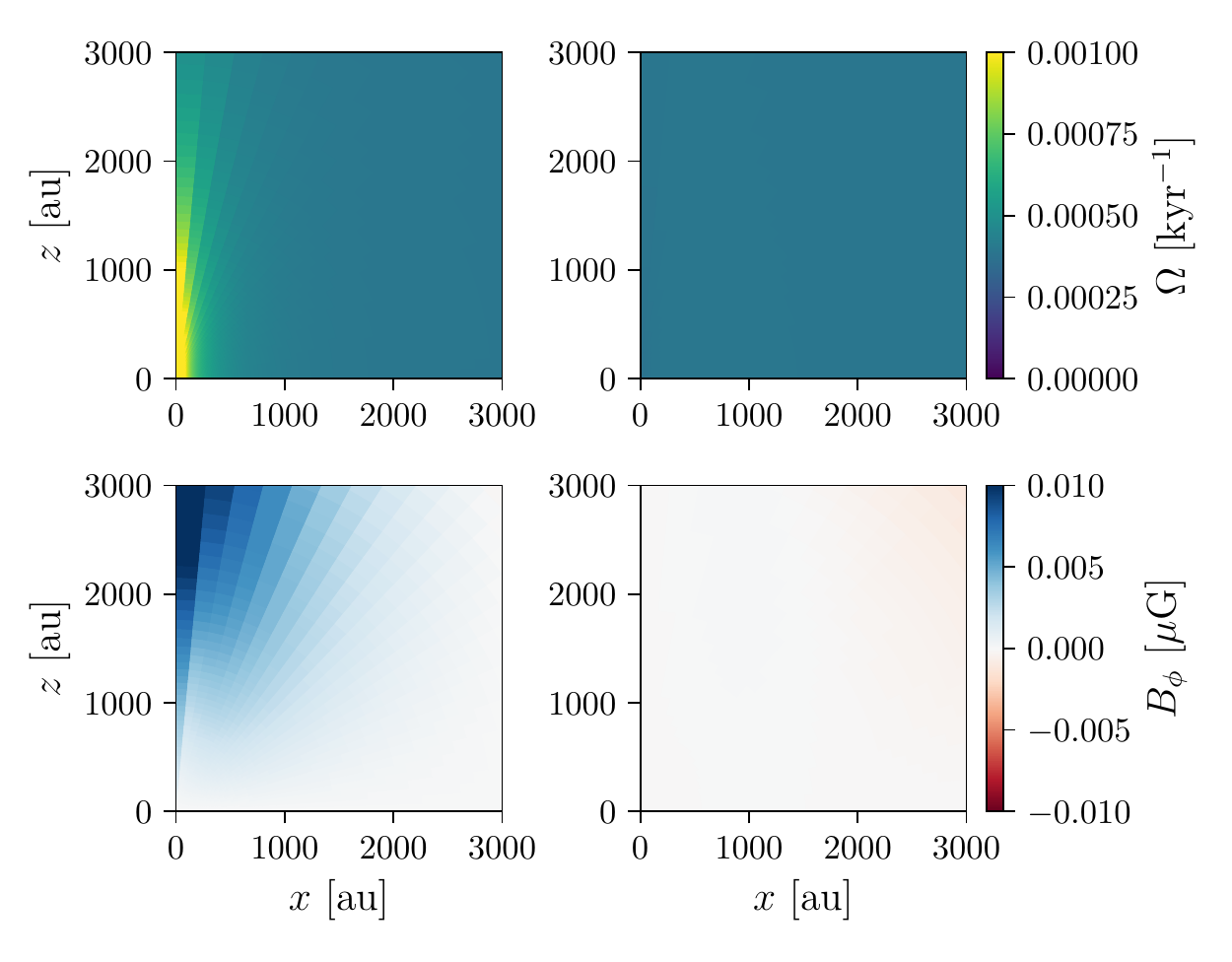}
    \vspace{-2em}
    \caption{Comparison between 2D simulations with (right) and without (left) the polar correction at $t=100~{\rm kyr}$. Physically, the inner few 1000~au shown here should still be rotating approximately uniformly. Without the polar correction, positive $B_\phi$ is generated unphysically; this leads to spin-up of the polar region. This problem can be fixed (in 2D) by introducing the polar correction discussed in Appendix \ref{A:polar_EMF}.}
    \label{fig:polar_comparison_2}
\end{figure}

During the first few 100 kyr of pre-stellar collapse, the evolution is slow and we expect the rotation in the inner part of the core to be approximately uniform (similar to the initial condition). However, as the top left panel of Fig.~\ref{fig:polar_comparison_2} shows, our simulations often show a faster rotating region near the pole, where the rotation rate is well above the initial rotation. This spin-up is unphysical, and the resulting unphysical increment of angular momentum in the innermost few $100~{\rm au}$ is likely tied to the relatively slow convergence with respect to $\theta$ resolution shown in Fig.~\ref{fig:ML_convergence}.

This problem is related to geometric truncation errors in the EMF calculation, which eventually lead to unphysical generation of a toroidal magnetic field (Fig.~\ref{fig:polar_comparison_2} bottom left panel) and unphysical angular-momentum transport.
The origin of this problem can be illustrated by considering a uniformly rotating system threaded by a vertical magnetic field. Physically, one expects the path integral of the poloidal EMF around a cell to be zero, so that the toroidal field remains zero. However, in a simulation with finite resolution, the EMF on each edge are subject to different geometric errors, and these errors in general do not cancel out. Additionally, in a spherical polar grid the amplitudes of such errors do not decrease quickly as resolution increases because the error for the first couple of cells around the pole is always ${\sim}\mathcal{O}(1)$.

\subsection{Implementing a polar correction}
\subsubsection{2D axisymmetric}

To avoid the unphysical spin-up observed in Fig.~\ref{fig:polar_comparison_2}, we introduce a correction in our code to ensure that a uniformly rotating system remains uniformly rotating, with no unphysical generation of $B_\phi$ or artifact in $\Omega$. This can be achieved, e.g., by modifying how the left and right interface values of $v_\phi$ are interpolated in the Riemann solver. (This correction is not applied to simulations shown in the main text.)

In {\tt Athena++}, for a 2D axisymmetric spherical polar grid, the poloidal EMF on cell edges (the path integral of which is used to update the toroidal field) is calculated using a Riemann solver on the cell interface in $r$ and $\theta$ directions.
For each interface, the code first interpolates cell-centered quantities to get left and right states on the interface, and then uses these left and right interface states to compute fluxes on the interface, including the face-centered EMF. Note that, for the $r$ and $\theta$ directions, the face-centered EMF is the same as the edge-centered EMF because of axisymmetry. The problem lies in how the azimuthal velocity $v_\phi$ is interpolated: the interpolation algorithm in {\tt Athena++} directly interpolates $v_\phi$ and, for a uniformly rotating system, the interpolated state on the interface sometimes differs from the actual $v_\phi$. This kind of error generally will not cancel out in the path integral of the EMF, making it possible to generate unphysical $B_\phi$.

Our fix to the problem is fairly simple. Instead of directly interpolating $v_\phi$, we first calculate cell-centered $\Omega=v_\phi/(r\sin\theta)$ and then interpolate $\Omega$ to the cell interface. 
We then compute the edge-averaged $v_\phi$ on the interface based on the interpolated $\Omega$, and use that as input for the Riemann solver. (Here, `edge-averaged' means averaging along an edge in the $r$ direction for a $\theta$ interface and along the $\theta$ direction for an $r$ interface.)
This eliminates geometric error when the system is uniformly rotating. This correction works very well for uniformly rotating systems, and applying it to our 2D simulations removes the artifact near the pole (Fig.~\ref{fig:polar_comparison_2}, right panels) without introducing any new artifacts.

\subsubsection{3D}

In 3D the correction above can no longer fully solve the problem. This is because the $r,\theta$ edge-centered EMFs are not the same as the corresponding interface EMFs in the absence of axisymmetry. Instead, now an edge EMF depends on all four adjacent cells and interfaces and that introduces additional geometric errors that are harder to account for.

\begin{figure}
    \centering
    \includegraphics[width=.4\textwidth]{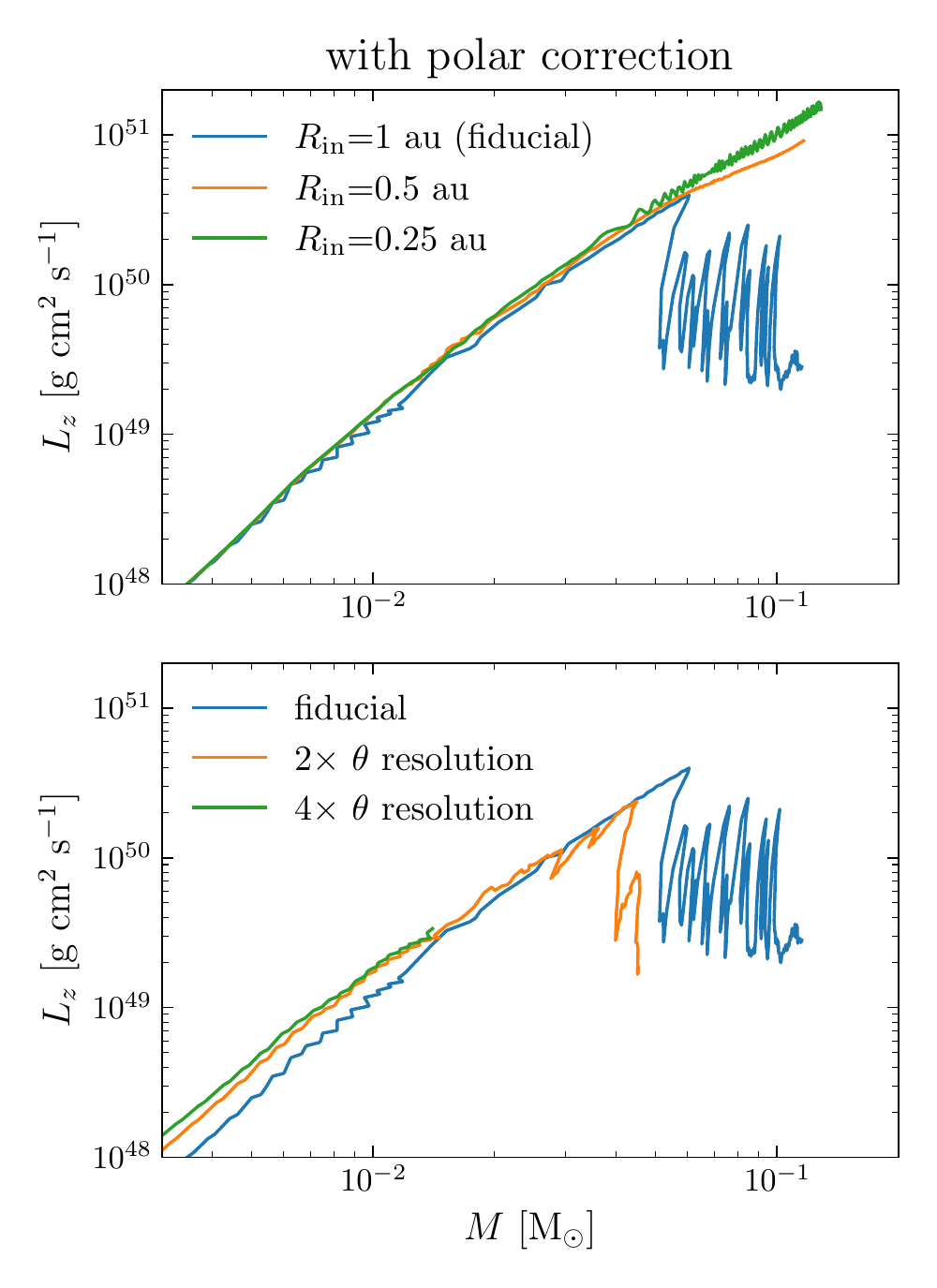}
    \caption{Mass--angular-momentum relation in 2D simulations with the polar correction. Better convergence in $\theta$ resolution than shown in Fig.~\ref{fig:ML_convergence} is achieved. Note that the drop in angular momentum at larger $M$ for runs with $r_{\rm in}=1~{\rm au}$ comes from mass ejection, which happens because the dense torus undergoes a violent instability when its size is very similar to the inner boundary. (The tori in these simulations are smaller than those in the main text, because the polar correction lowers their angular momentum.)}
    \label{fig:ML_corrected}
\end{figure}

\begin{figure}
    \centering
    \includegraphics[width=.4\textwidth]{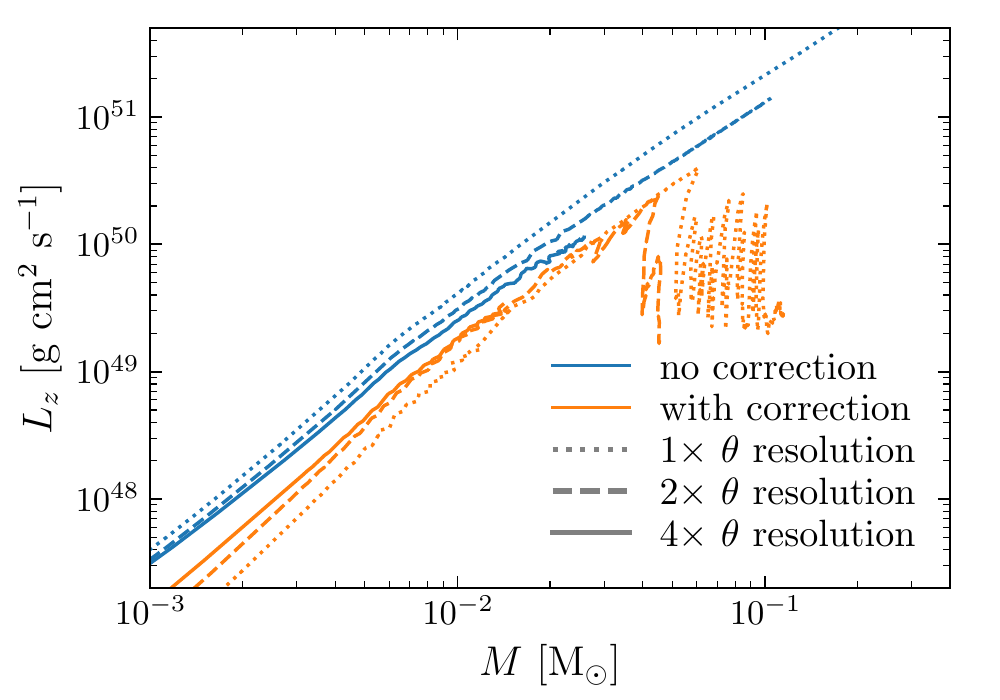}
    \caption{Mass--angular-momentum relation in 2D simulations with and without the polar correction at different $\theta$ resolutions. All runs use $r_{\rm in}=1~{\rm au}$. The two different treatments converge towards the same result.}
    \label{fig:ML_corrected_comparison}
\end{figure}

\begin{figure}
    \centering
    \includegraphics[width=.5\textwidth]{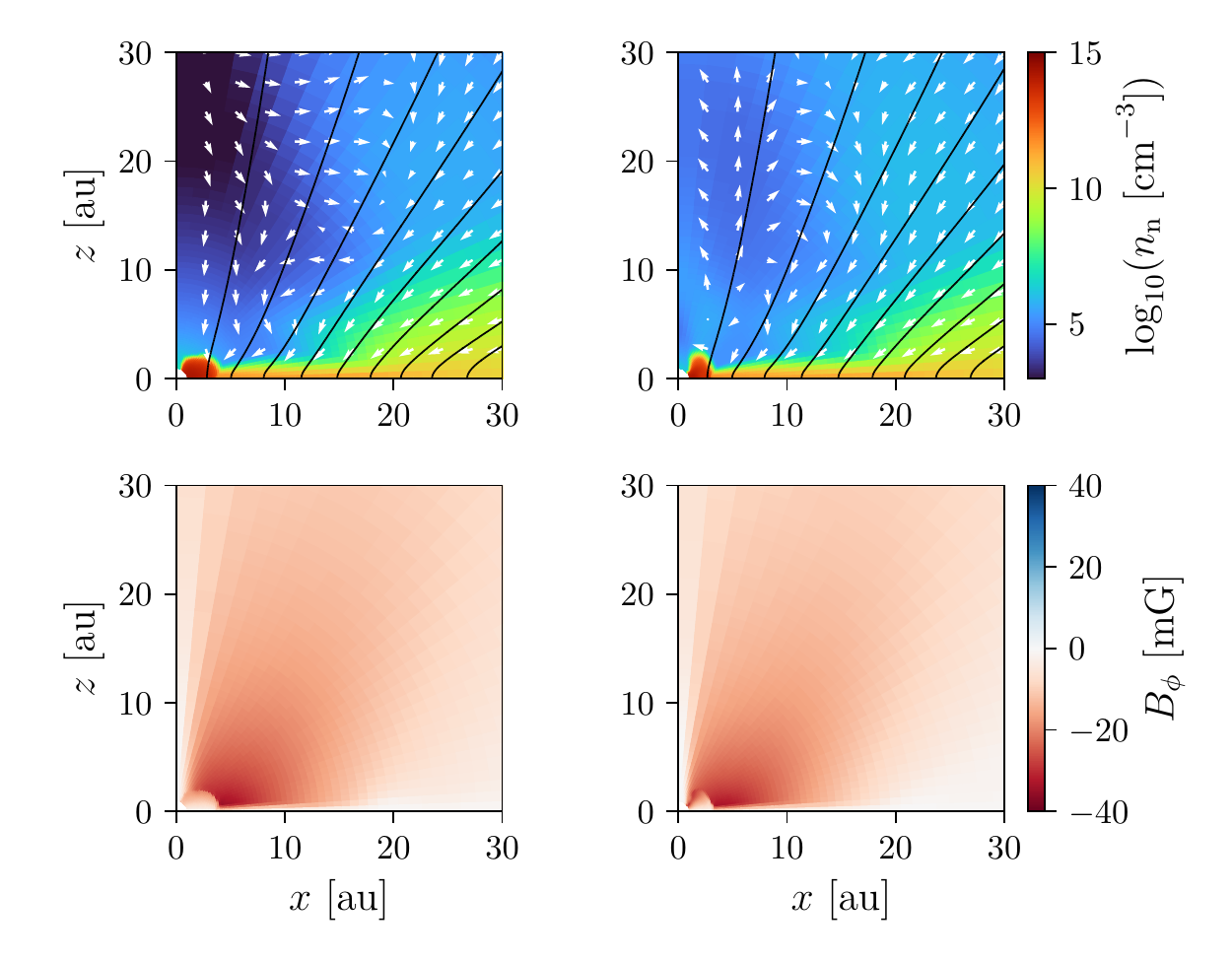}
    \vspace{-2em}
    \caption{Comparison between 2D simulations with (right) and without (left) the polar correction at $M_{10}\approx 0.05~{\rm M}_\odot$. Unlike at earlier times, the behaviours are qualitatively very similar.}
    \label{fig:polar_comparison}
\end{figure}

\subsection{Comparison between simulations with and without polar correction}

To check whether the unphysical spin-up we observed has a significant impact on our physical results regarding disc formation, we rerun most of our 2D simulations with the polar correction discussed above and compare the results. We find the qualitative evolution to be overall similar, and simulations with polar correction show good numerical convergence with respect to inner-boundary size and resolution (Fig.~\ref{fig:ML_corrected}).
Also, comparing simulations with and without polar correction, we find that the mass--angular-momentum relation converges to the same limit from two different directions as resolution increases (Fig.~\ref{fig:ML_corrected_comparison}).
Assuming that simulations with and without the polar correction both converge to the true mass--angular-momentum relation, we conclude that our fiducial resolution simulations without polar correction overestimate the angular momentum by roughly a factor of 2 at late times (after the formation of protostar).

It is also worth noting that, although the unphysical spin-up significantly affects the evolution of the innermost few ${\sim}100~{\rm au}$ at early times (Fig.~\ref{fig:polar_comparison_2}), the qualitative evolution after the formation of the protostar is much less affected, as is shown in Fig.~\ref{fig:polar_comparison}.

\section{A note on computational cost}
\label{A:cost}

As we mentioned in Section \ref{subsec:intro_sims}, a main challenge for long-term simulations of protostellar disc formation is the computational cost, or more precisely the timestep. Here we discuss how we tackle this problem.

Typically, the numerical timestep is limited by magnetic diffusion at the inner edge of the disc. One approach to overcoming this obstacle is to increase the cell size there. This in turn requires increasing the inner boundary size, since vertically resolving the disc already poses a requirement on angular resolution.
As we show in \S\ref{sec:convergence}, this can be achieved without hurting numerical convergence by treating the inner boundary carefully. For example, the maximum $r_{\rm in}$ one can take while maintaining numerical convergence differs by at least a factor of four between a simple open boundary condition and our boundary condition, and this can speed up simulations by above an order of magnitude. This allows numerically converged long-term 3D simulations to be performed, for our code, on ${\sim}100$ processors over a timescale of several weeks.

Another option to reducing the computational cost is to perform 2D simulations, while using parametrized sub-grid models to account for the effects of non-axisymmetric structure (e.g., spiral waves induced by gravitational instability, which transport angular momentum). We plan to model the effects of gravitational self-regulation in this fashion in a future paper. If such modelling is possible (and gives results in good agreement with 3D simulations), this will be a particularly useful tool for performing large parameter surveys for which each long-term 2D simulation would only cost a few days on a few tens of cores.
\label{lastpage}

\end{document}